\documentclass[twocolumn,
 superscriptaddress,
 amsmath,amssymb,
 aps,
 floatfix
]{revtex4-2}

\usepackage{amsmath, amssymb, amsthm}
\usepackage[usenames,dvipsnames]{xcolor}
\usepackage{graphicx}
\usepackage{dcolumn}
\usepackage{bm}
\usepackage{mathrsfs}
\usepackage{hyperref}
\usepackage[capitalise]{cleveref}

\usepackage{wrapfig}
\usepackage{mathrsfs}

\newtheorem{proposition}{Proposition}[section]
\newtheorem{lemma}[proposition]{Lemma}

\DeclareMathOperator{\sech}{sech}

\newcommand{\eexp}[1]{e^{#1}}
\newcommand{\norm}[1]{\left\lVert#1\right\rVert}

\begin{document}

\title{Free Entropy Minimizing Persuasion in a Predictor-Corrector Dynamic\footnote{This is a draft document. Results are subject to revision. This document has not been peer-reviewed.}}

\author{Geoff Goehle}
\email{goehle@psu.edu}
\author{Christopher Griffin}%
\email{griffinch@psu.edu}
\affiliation{
	Applied Research Laboratory,
    The Pennsylvania State University,
    University Park, PA 16802
    }%


\begin{abstract} Persuasion is the process of changing an agent's belief distribution from a given (or estimated) prior to a desired posterior. A common assumption in the acceptance of information or misinformation as fact is that the (mis)information must be consistent with or familiar to the individual who accepts it. We model the process as a control problem in which the state is given by a (time-varying) belief distribution following a predictor-corrector dynamic. Persuasion is modeled as the corrector control signal with the performance index defined using the Fisher-Rao information metric, reflecting a fundamental cost associated to altering the agent's belief distribution. To compensate for the fact that information production arises naturally from the predictor dynamic (i.e., expected beliefs change) we modify the Fisher-Rao metric to account just for information generated by the control signal. The resulting optimal control problem produces non-geodesic paths through distribution space that are compared to the geodesic paths found using the standard free entropy minimizing Fisher metric in several example belief models: a Kalman Filter, a Boltzmann distribution and a joint Kalman/Boltzmann belief system.
\end{abstract}

\maketitle

\section{Introduction}
Persuasion has been studied extensively (see \cite{O15} for example). Despite its long history (most likely as long as language has been present), there are fewer mathematical models of persuasion. Two early models of persuasion were put forth by Cervin and Henderson \cite{CH61} whose models were purely statistical. Burgoon et al. \cite{BBMS81} relate persuasion to learning theory, which we exploit as part of our justification for an action to be minimized, but do not provide a formal mathematical model. More recently, Curtis and Smith \cite{CS08} study persuasion dynamics using a model that is classicaly used in opinion dynamics \cite{DeGroot74,Krause00,HK02,BN05,WDA05,Toscani06,Weisb06,Lorenz07,BHT09,CFL09,KR11,DMPW12,CFT12,JM14,SZAS21,GG21}, which can be thought of as a sub-discipline of consensus or flocking theory \cite{DeGroot74,Krause00,Centola15,TT98,CS07,EK01,L08,LX10, DM11,MT14,GSJ22}. Similarly, Huang et al. \cite{HZXF16} study consensus on complex networks with persuasion, using a threshold model that is similar (in spirit) to a diffusion decision model \cite{RM08} used in cognitive neuroscience. Persuasion in product ratings on networks has been studied by Xie et al. \cite{XZLL21} using Markov chains to study cascading phenomena. So-called Bayesian persuasion has been investigated in the economics literature beginning with Kamenica and Gentzkow \cite{KG11} who phrased the problem in terms of mechanism design. This work was extended by Babichenko \cite{BTZ21} to include constraints. Most closely related to this work is that of Caballero and Lunday \cite{CL19} who present an expected utility maximization framework, constructing a bilevel optimization problem to describe and optimize persuasion action.

The related area of misinformation has been studied extensively over the past decades \cite{ABA23,DBZP16,VB20}. Consequently, the references presented cannot cover the entire breadth of the field. Much of this research focuses on political misinformation \cite{E01,JZ20}, health misinformation \cite{SLo20,SNCG19} (especially related to COVID-19 \cite{RSDK20,JFKE22,GBFR22} or anti-vaccine conspiracies \cite{WFCL23,NEER22,EUKS22}) and climate change \cite{C22,ZS22,FM23}. Countering misinformation is also widely studied \cite{CEL15,VMCL17,VLRM17,THKE22,FM23}, with limited consensus \cite{ESS23} on effective mitigation strategies except for misinformation vaccination \cite{SB22,BSSS22,VMCL17}. Technological solutions focus on the development of robust classifiers for a continuously changing misinformation landscape \cite{FM22,SSZJ22,FGLP23,MKSM22}. However, a fundamental question that is still under investigation is why individuals are susceptible to misinformation in the first place. Most research on understanding the acceptance of misinformation relies on six tenets of cognitive psychology \cite{ELCS22}:
\begin{enumerate}
\item Intuitive thinking, including a lack of analysis.
\item Cognitive failure, including ignoring or forgetting source cues or prior (counter) knowledge.
\item Illusory truth, including familiarity or consistency of the false information.
\item Social cues, including presentation by elites or in-group sources.
\item Emotional state of the receiver.
\item World view, reflecting personal views or partisanship.
\end{enumerate}
While it is too much to assume a simple mathematical model will capture all of these elements (especially, e.g., the emotional state of an individual), tenets 3 and 6 are consistent with the hypothesis that individuals are more susceptible to misinformation or persuasion if the presented information is consistent with their pre-existing mental models (see also \cite{R20,WT20,SCM20,ELCS22, C22a}, for further support). Viewed as an information coding problem \cite{BM14}, it is sensible to apply an information theoretic approach to capturing the notion of consistency within a mathematical model of persuasion and misinformation.

The main contributions of this paper are:
\begin{enumerate}
\item We formulate a simple model of persuasion that can incorporate misinformation using a dynamic predictor-corrector framework. In this framework, information injection is modeled as the corrector.

\item We introduce the Fisher-Rao metric as the natural metric on the space of possible belief distributions since it can be used to construct free entropy geodesics, i.e., paths that minimize the incremental Kullbach-Leibler divergence between distributions.

\item We show how to alter the action corresponding to free entropy \cite{C07} to account for the (accepted) knowledge of distribution change as a result of the predictor. The resulting optimization problem is then an optimal control problem rather than a problem in the calculus of variations.

\item For example belief dynamics, we compare solutions from the uncorrected action minimization problems to solutions of the optimal control problems and show how to construct a fusion model over multiple distributions.
\end{enumerate}

The remainder of this paper is organized as follows: We provide mathematical preliminaries in \cref{sec:Prelims}. In \cref{sec:GeneralModel}, we discuss our general predictor-corrector model and the resulting optimal control problem. Results on specific predictor-corrector dynamics and their relation to neuroscience are presented in \cref{sec:Kalman,sec:Boltzmann,sec:Fusion}. Conclusions and future directions are presented in \cref{sec:Future}. \cref{app:OC} contains a brief overview of optimal control results.

\section{Mathematical Preliminaries}\label{sec:Prelims}
Techniques from information geometry have been used extensively in statistical physics \cite{B11,SSBC12,SC12,KLHH16,FC09,KH17,K21,GPB20,NGCG20}, especially for studying far from equilibrium processes and phase transitions. These techniques have also found application in dynamical systems and population dynamics \cite{FA95,ZGS20,PE13} as well as the dynamics of neural networks \cite{A97} and machine learning \cite{OAAH17}. In this section, we provide a brief overview of key results in information geometry used in our analysis.

Let $\Omega$ be a sample space and let $\mathscr{P}(\Omega;\bm{\theta})$ be the set of all probability distributions over this sample space that are parametrized by a vector $\bm{\theta} = (\theta^1,\dots,\theta^n)$. If $P(\bm{\theta}) \in \mathscr{P}(\Omega;\bm{\theta})$, let $p(\mathbf{x}|\bm{\theta})$ be the corresponding distribution function. The Kullback-Leibler (KL) divergence \cite{Kullback1951} is an entropy based statistical distance that does not satisfy the conditions required to be a true metric. For distributions $P$ and $Q$ with density functions $p$ and $q$, it is defined as
\begin{equation}
D_{KL}(P\|Q) = \int_X p(x) \ln\left(\frac{p(x)}{q(x)}\right) dx.
\end{equation}
(See \cite{bishop2006} for details.) From this definition, it is straightforward to see that the KL-divergence is neither symmetric nor does it satisfy any version of the triangle inequality. The value of the KL-divergence has units of ``nats'', or, if $\log_2$ is used instead, has units of ``bits''. Informally, the KL-divergence measures the amount of excess information (i.e., Shannon information or ``surprise'') resulting from using $Q$ as a model for data generated by $P$.

The Fisher-Rao pseudo-Riemannian metric fully generalizes the KL-divergence, in the sense that its units are in terms of information (nats) and by Chentsov's theorem it is the only Riemannian metric invariant under sufficient statistics \cite{D18}. The quadratic form of the metric is given by
\begin{equation}
\mathbf{g}_{jk}(\bm{\theta}) = \int_X \frac{\partial\log[p(\mathbf{x}|\bm{\theta})]}{\partial \theta^j} \frac{\partial\log[p(\mathbf{x}|\bm{\theta})]}{\partial \theta^k} p(\mathbf{x}|\bm{\theta}) dx.
\label{eqn:QuadraticForm}
\end{equation}
This quadratic form can be defined directly in terms of the Hessian of the self-information of the distribution or in relation to the partition function if the distribution is derived from a Gibbs measure \cite{C15}.

To define a geodesic, we use the kinetic energy action, where we make use of Einstein summation notation here and in the following,
\begin{equation}
A = \frac{1}{2} \int_{t_0}^{t_f}  \mathbf{g}_{jk}(\bm{\theta}) \frac{\partial\theta^j}{\partial t} \frac{\partial\theta^k}{\partial t} \,dt.
\label{eqn:entact}
\end{equation}
A geodesic is then any path $\bm{\theta}(t)$ with $t\in[t_0,t_f]$ that minimizes the defined action. This perspective is particularly important in statistical mechanics \cite{C07}, as minimizing the change in free entropy can reduce the amount of energy needed to transform an ensemble.  In terms of information geometry \cref{eqn:entact} represents the total change in information as the system evolves from $t_0$ to $t_f$ and would have units of nats.  It follows from the Cauchy-Schwartz inequality that \cref{eqn:entact} has a minimum when one follows the geodesics defined by the metric.

In general geodesics for the Fisher-Rao metric may be difficult to compute.  There are specific situations, see below, where the geodesics can be constructed.  However it is not necessary to solve for the geodesics in general if we are given specific boundary values.  Instead we use the Euler-Lagrange equations. Let
\begin{equation*}
\mathcal{L(}t, \bm{\theta}, \dot{\bm{\theta}}) = \mathbf{g}_{jk}(\bm{\theta}) \dot{\theta}^j \dot{\theta}^k,
\end{equation*}
then the geodesics are solutions to the Euler-Lagrange equations with appropriate boundary conditions on $\bm{\theta}$,
\begin{equation}
\label{eqn:euler-lagrange}
\frac{\partial \mathcal{L}}{\partial \theta^i} - \frac{d}{dt}\frac{\partial \mathcal{L}}{\partial \dot{\theta}^i} = 0.
\end{equation}
In \cref{sec:GeneralModel}, we will generalize the action so that the classical calculus of variations no longer applies and our Euler-Lagrange equations will be replaced by those from modern optimal control theory \cite{K04}, see Appendix~\ref{app:OC}.

The Fisher-Rao metric can be interpreted by its relationship to KL-divergence.  Consider two points in the parameter space $\bm{\theta}_0$ and $\bm{\theta}$ that are infinitesimally close (in the Euclidean sense).  Then it can be shown \cite{cover2005} that the second order Taylor expansion of the map $\bm{\theta}_0 \mapsto D_{KL}[P(\bm{\theta})\|P(\bm{\theta}_0)]$ is given by
\begin{equation}
D_{KL}\left[P(\bm{\theta})\|P(\bm{\theta}_0)\right] = \frac{1}{2} \mathbf{g}_{jk}(\bm{\theta}_0)\Delta\theta_0^j\Delta\theta_0^k + O(\norm{\bm{\Delta\theta}_0}^3).
\label{eqn:fisherkl}
\end{equation}
The implication of this relationship is that at close distances the Fisher-Rao metric is defined (in some sense) by the KL-divergence and is measuring a change in information as the distribution parameters are perturbed. Indeed, if we choose a small time step $\Delta t$ and divide a path from $\bm{\theta}(t_0)$ to $\bm{\theta}(t_f)$ so that at $i\Delta t$ we move from $\bm{\theta}_{i-1}$ to $\bm{\theta}_i$ we can use \cref{eqn:fisherkl} to approximate the entropy action \cref{eqn:entact} by,
\begin{multline}
A \approx \sum_i \frac{1}{2} \mathbf{g}_{jk}(\bm{\theta}_i)\frac{\partial\theta^j}{\partial t}  \frac{\partial\theta^k}{\partial t} \Delta t \approx \\
\frac{1}{\Delta t} \sum_i \frac{1}{2} \mathbf{g}_{jk}(\bm{\theta}_i)\Delta\theta_i^j\Delta\theta_i^k  \approx \\
\frac{1}{\Delta t} \sum_i D_{KL}\left[P(\bm{\theta}_{i+1}) \| P(\bm{\theta}_{i})\right].
\label{eqn:klapprox}
\end{multline}
The appropriate interpretation of \cref{eqn:klapprox} is that if an observer were to periodically sample the changing distribution and compute the KL-divergence between time steps, the value of the action would represent (a time-step normalized version of) the total Shannon information (i.e., surprise at traveling along the parameter path).

\begin{figure}
\includegraphics[width=0.9\columnwidth]{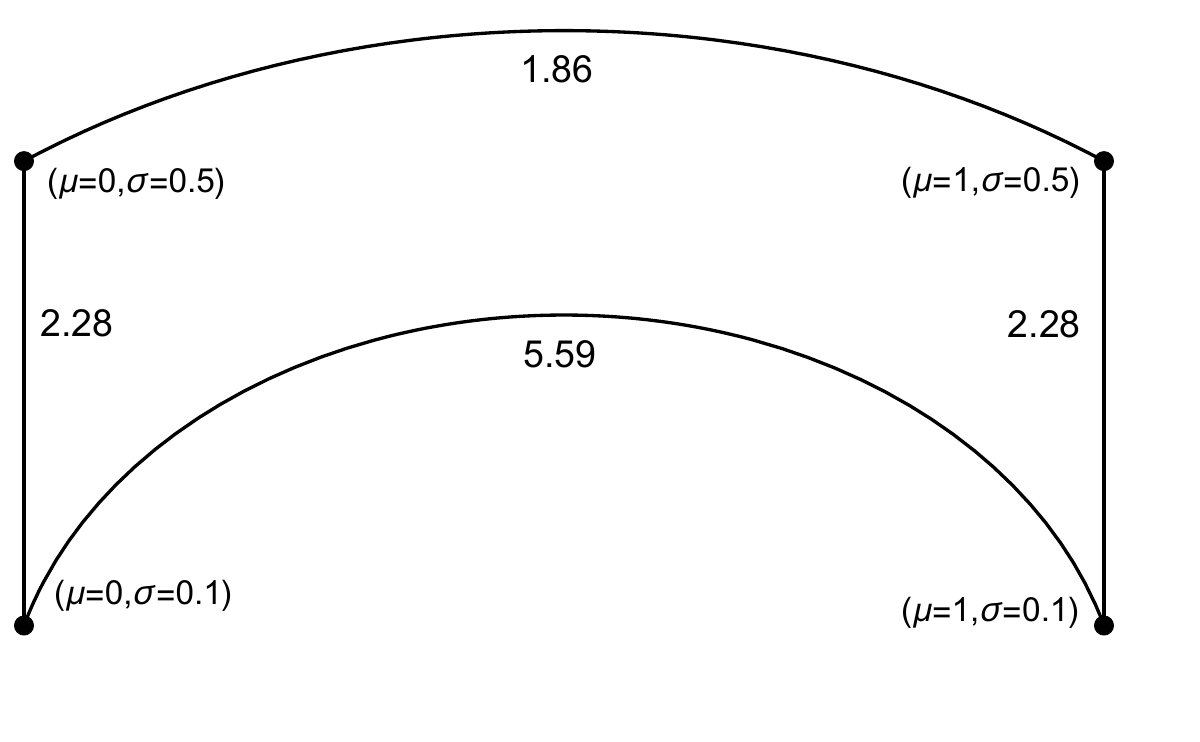}
\caption{Four normal distributions in the Poincar\'e half-plane model.  Curves represent geodesics and are labeled by the Fisher-Rao distance.}
\label{fig:halfplane-cartoon}
\end{figure}

By way of example, and to establish results that will be used later, we briefly present a classic result from the information geometry of the Gaussian distribution,
\begin{equation*}
p(x|\mu,\sigma) = \frac{1}{\sqrt{2\pi\sigma^2}}\exp\left[-\frac{(x-\mu)^2}{2\sigma^2}\right].
\end{equation*}
Using \cref{eqn:QuadraticForm}, the metric can be computed in \((\mu,\sigma)\) order as
\begin{equation*}
\mathbf{g} = \frac{1}{\sigma^2}\begin{bmatrix}
1 & 0 \\
0 & 2
\end{bmatrix}.
\end{equation*}
This is a minor variation on the metric of the Poincar\'{e} half-plane and thus the space of Gaussian distributions exhibits hyperbolic geometry. Minimizing the action from \cref{eqn:entact} leads to the Euler-Lagrange equations
\begin{align*}
&\ddot{\mu} = \frac{2\dot{\mu}\dot{\sigma}}{\sigma}\\
&\ddot{\sigma} = \frac{2\dot{\sigma}^2 - \dot{\mu}^2}{2\sigma},
\end{align*}
which have solutions,
\begin{align}
\mu = \eta + 2\rho\tanh[\omega(t-\kappa)]\label{eqn:muopt}\\
\sigma = \sqrt{2}\rho\sech[\omega(t-\kappa)],\label{eqn:sigmaopt}
\end{align}
where the constants $\eta$, $\rho$, $\kappa$ and $\omega$ are determined from the initial and final conditions on the system of ODEs. From hyperbolic trigonometric identities, it follows at once that
\begin{equation*}
\left(\frac{\mu - \eta}{2\rho}\right)^2 + \left(\frac{\sigma}{\sqrt{2}\rho}\right)^2 = 1,
\end{equation*}
and we see that solutions in phase space must trace out arcs on the $\sigma$-positive branch of an ellipse centered at $(\eta,0)$, as expected (see \cref{fig:halfplane-cartoon}).

\section{General Model}\label{sec:GeneralModel}
We use a simplified form of the Bayesian brain hypothesis \cite{RB99,F12,BI21,MTSB21,XCBL13} to model an individual's belief using a linear predictor-corrector system
\begin{equation}
\dot{\bm{\theta}} = \bm{\phi}(\bm{\theta}, t) + \bm{\chi}(\bm{\theta}, \bm{\eta}, t).
\label{eqn:theta-model}
\end{equation}
Here $\bm{\theta}$ are the parameters of the belief distribution $P(\bm{\theta})$ (to be influenced), $\bm{\phi}$ represents the agent's prediction of how the belief distribution changes over time, and $\bm{\chi}$ represents corrections resulting from the integration of evidence (information) $\bm{\eta}$. While we will not do so here, this linearization could be refined through the addition of $\bm{\phi}\times\bm{\chi}$ cross-terms.

An effective persuasion signal $\bm{\eta}$ will drive the parameters from some initial condition $\bm{\theta}_0$ to a final condition $\bm{\theta}_f$. The path traversed through probability space has free entropy given by \cref{eqn:entact}. Consequently, the optimal control problem that minimizes the free entropy is given by
\begin{equation}
\begin{aligned}
\min \;\; & \frac{1}{2} \int_{t_0}^{t_f}  \mathbf{g}_{jk}(\bm{\theta}) \frac{\partial\theta^j}{\partial t} \frac{\partial\theta^k}{\partial t} \,dt\\
s.t. \;\; & \dot{\bm{\theta}} = \bm{\phi}(\bm{\theta}, t) + \bm{\chi}(\bm{\theta},\bm{\eta},t)\\
& \bm{\theta}(t_0) = \bm{\theta}_0 \quad \bm{\theta}(t_f) = \bm{\theta}_f\\
&\bm{\eta}\in L^2[\mathbb{R}^k].
\end{aligned}
\label{eqn:ProbA}
\end{equation}
However, the action in this case unnecessarily penalizes the natural evolution of the belief as a result of the uncorrected dynamics $\dot{\bm{\theta}} = \bm{\phi}(\bm{\theta}, t)$. That is, it is penalizing information that would be expected by a Bayesian brain and therefore already consistent with the belief model.

To model the true cost of persuasion in terms of the inconsistency of inputs with an individual's naturally time-varying belief function, we must define an action that  does not penalize the natural time-evolution of the belief surface. As such, we will use the modified action
\begin{equation}
A_\chi = \frac{1}{2}  \int_{t_0}^{t_1} \mathbf{g}_{jk}\chi^j \chi^k dt,
\label{eqn:corraction}
\end{equation}
in \cref{eqn:ProbA}.

As with \cref{eqn:entact}, this action has an information theoretic interpretation. It follows from \cref{eqn:theta-model} that
\begin{equation*}
\bm{\theta}(t_{i+1}) \approx \bm{\phi}\Delta t + \bm{\chi}\Delta t.
\end{equation*}
Our requirement that the action does not penalize the natural time-evolution of the belief surface corresponds to disregarding changes in $\bm{\theta}$ due to the prediction term $\bm{\phi}$, which represents prior dynamics in the belief process, and only accumulating information from the corrector term $\bm{\chi}$. In other words we wish to minimize the average KL-divergence between the predicted and corrected belief distributions at time $t_i$ given by,
\begin{equation*}
D_{KL}\left[P(\bm{\theta}(t_i)+\bm{\phi}\Delta t) \| P(\bm{\theta}(t_{i})+\bm{\phi}\Delta t+\bm{\chi}\Delta t)\right].
\end{equation*}
Using \cref{eqn:fisherkl} we have
\begin{multline*}
A_\chi \approx \sum_i \mathbf{g}_{jk}[\bm{\theta}(t_i)] \chi^i \chi^k \Delta t \approx \\
\frac{1}{\Delta t} \sum_i \mathbf{g}_{jk}[\bm{\theta}(t_i) + \bm{\phi}\Delta t] \chi^i\Delta t\chi^k\Delta t \approx \\
\frac{1}{\Delta t} \sum_i D_{KL}\left[P(\bm{\theta}(t_i)+\bm{\phi}\Delta t+\bm{\chi}\Delta t) \| P(\bm{\theta}(t_{i})+\bm{\phi}\Delta t)\right].
\end{multline*}
Thus \cref{eqn:corraction} measures accumulation of information due to dynamics resulting from the corrector term.  In situations where $\bm{\chi} = \bm{0}$ the system will evolve according to its natural predictor dynamics and $A_\chi = 0$.  In situations where $\bm{\phi} = 0$, then $A_\chi = A$ and we recapture the geodesic interpretation presented previously.

Whether explicit or intuitive, observers often have a threshold for the rate at which they expect to gain information. Aggressive correctors, which cause belief distributions to change quickly so that $A_\chi$ is large, are likely to be viewed with suspicion, based on empirical evidence \cite{ELCS22}.  On the other extreme, if the KL-divergence of an individual step is small enough it may be difficult for the observer to detect the impact of the corrector at all.  By optimizing \cref{eqn:corraction} to achieve an optimal path we can, notionally, change the underlying distribution slowly enough (with respect to the information metric) that the observer is less likely to reject persuasion (or more likely to be susceptible to misinformation).

For the remainder of this paper, we will refer to the optimization problem with action given in \cref{eqn:entact} as the {\em geodesic control problem}. The optimization problem with action given by \cref{eqn:corraction} will be referred to as the {\em corrector control problem}.


\section{Persuasion in the Kalman-Bucy Filter}\label{sec:Kalman}
Kalman filters are archetypal examples of Bayesian filters \cite{MS83} and as such can be used to model belief/decision processes in neurological systems \cite{XCBL13}, under a Bayesian brain assumption. More generally, they are also a well established model for perception, particularly in control theory (see e.g., \cite{R99}).

Suppose that an object is moving in one dimension with constant velocity $u$ so that
\begin{align*}
&\dot{x} = u + w\\
&z = x + v,
\end{align*}
where $z$ is the observation and $w \sim N(0,q)$ and $v \sim N(0,r)$. Here $q$ and $r$ are variances. The associated Kalman-Bucy filter \cite{lewis2008} produces a distribution estimate $N(\mu, s)$, with
\begin{align}
&\dot{\mu} = u + \frac{s}{r}(z-\mu)\label{eqn:mudot}\\
&\dot{s} = q - \frac{s^2}{r},\label{eqn:sdot}
\end{align}
where $s = \sigma^2$ is again a variance. It is straightforward to see how these equations can be written in predictor-corrector form as,
\begin{equation*}
\begin{bmatrix}\dot{\mu}\\ \dot{s}\end{bmatrix} =
\underbrace{\begin{bmatrix}u\\q\end{bmatrix}}_{\bm{\phi}} +
\underbrace{\begin{bmatrix}\frac{s}{r}(z-\mu)\\-\frac{s^2}{r}\end{bmatrix}}_{\bm{\chi}}.
\end{equation*}
In this problem, the control variables are,
\begin{equation*}
\bm{\eta} = \begin{bmatrix}z\\r\end{bmatrix},
\end{equation*}
since they characterize the observed (input) signal. The persuasive problem in this case is to provide inputs (real or fictitious) to drive the state of belief to some $(\mu_f,s_f)$, given an initial belief state $(\mu_0,s_0)$. We now consider the problems of finding an optimal input $\bm{\eta}$ in both \cref{eqn:ProbA} and the alternate problem using the action given in \cref{eqn:corraction}.

\subsection{Geodesic Control Problem}

We can solve the geodesic control problem in \cref{eqn:ProbA} using \cref{eqn:muopt,eqn:sigmaopt}. Solving \cref{eqn:mudot} and \cref{eqn:sdot} in terms of $r$ and $z$ yields,
\begin{align*}
&z = \frac{\mu(q - \dot{s}) + s(\dot{\mu} -u)}{q - \dot{s}} = \mu + s \frac{\dot{\mu} -u}{q - \dot{s}}.\\
&r = \frac{s^2}{q - \dot{s}}.
\end{align*}
Here
\begin{equation*}
s = \sigma^2 = 2\rho^2\sech^2[\omega(t-\kappa)].
\end{equation*}
For this solution to be well behaved, we require
\begin{equation*}
q - \dot{s} = q+4 \rho ^2 \omega  \tanh (\omega  (t-\kappa )) \text{sech}^2(\omega  (t-\kappa )) > 0.
\end{equation*}
Therefore,
\begin{equation*}
q > \arg\min_{0\leq t \leq T} \left[4 \rho ^2 \omega  \tanh (\omega  (t-\kappa )) \text{sech}^2(\omega  (t-\kappa ))\right].
\end{equation*}
The critical points of $q - \dot{s}$ are at
\begin{equation*}
t_{1,2}^* = \kappa + \frac{\log(2\pm \sqrt{3})}{2\omega}.
\end{equation*}
At these points we have
\begin{equation*}
q-\dot{s}\left[\kappa + \frac{\log(2\pm \sqrt{3})}{2\omega}\right] = q \mp \frac{8 \rho ^2 \omega}{3 \sqrt{3}}.
\end{equation*}
Depending on the sign of $\omega$, the two critical points $t_{1,2}^*$ are a (local) maximum and minimum respectively. Assuming $\omega \neq 0$, a simple computation shows that
\begin{equation*}
\lim_{t \to \pm\infty} q - \dot{s}(t) = q.
\end{equation*}
Therefore, the critical points must correspond not just to local extrema but to a global maximum and minimum respectively. It follows that a necessary and sufficient condition for the solutions of $r$ and $z$ to be well-behaved (not experience blowup in finite time) is
\begin{equation}
\label{eqn:qcond}
q > \left\lvert \frac{8 \rho^2 \omega}{3 \sqrt{3}} \right\rvert.
\end{equation}
Notably, this inequality could also be reframed into a condition on the time horizon and is equivalent to
\begin{equation*}
(t_f - t_0) > \left(\frac{|\alpha|}{3\sqrt{3} \beta q}\right)^{1/3}
\end{equation*}
where \(\alpha\) and \(\beta\) are constants determined by the boundary conditions. Observe that the smaller $q$, the larger the minimum possible time horizon $t_f - t_0$.  Intuitively this can be interpreted as very strong priors result in a longer persuasion time.

\subsection{Corrector Control Problem}

In contrast, we consider the corrector control problem. Using \cref{eqn:corraction}, we obtain the action,
\begin{equation*}
A_\chi = \frac{s^2}{4 r^2}+\frac{s (z-\mu )^2}{2 r^2},
\end{equation*}
since the metric written in terms of $s$ (rather than $\sigma$) is,
\begin{equation*}
\mathbf{g} = \begin{bmatrix}
\frac{1}{s} & 0 \\
0 & \frac{1}{2s^2}
\end{bmatrix}.
\end{equation*}
The Hamiltonian for the control system is given as
\begin{equation}
H  = A_\chi   + \lambda\left(u + \frac{s}{r}(z-\mu)\right) + \nu\left(q - \frac{s^2}{r}\right),
\label{eqn:KalmanHamiltonianCorrector}
\end{equation}
where $\lambda$ and $\nu$ are Lagrange multipliers (the co-state variables).  Setting $H_\eta = 0$ yields solutions for the inputs,
\begin{align}
z &= \mu - \frac{\lambda}{2\nu} \label{eqn:z} \\
r &= \frac{1}{2\nu}. \label{eqn:r}
\end{align}
The resulting Euler-Lagrange equations show that $\dot{\lambda} = 0$ and consequently, we define the constant $\lambda(t) = \Lambda$. The remaining Euler-Lagrange equations are,
\begin{align}
\dot{\mu} &= u - \Lambda s \\
\dot{s} &= q - 2\nu s^2 \\
\dot{\nu} &= \frac{\Lambda^2}{2} + 2\nu^2 s.
\end{align}
Notice, $\dot{s}$ and $\dot{\nu}$ are independent of $\mu$ and form a (classical) Hamiltonian system with first integral (mechanical Hamiltonian),
\begin{equation}
\mathcal{H} = \frac{\Lambda^2s}{2} - q\nu + \nu^2s^2.
\label{eqn:HKF}
\end{equation}
Here $\mathcal{H}$ is a constant implicitly determined by the boundary conditions (as is $\Lambda$).

While open-loop controls (i.e., functions $z(t)$ and $r(t)$) do not appear to be easily constructed, it is possible to use this information to obtain closed loop controls $z(\mu,s)$ and $r(\mu,s)$. From the \cref{eqn:z,eqn:r},
\begin{equation*}
z = \mu - \Lambda r.
\end{equation*}
We can use \cref{eqn:HKF} to construct solutions for $r$ as,
\begin{equation*}
r = \frac{-q\pm \sqrt{q^2 + 2s^2\zeta(s)}}{2\zeta(s)},
\end{equation*}
where
\begin{equation*}
\zeta(s) = 2\mathcal{H} - s\Lambda^2.
\end{equation*}
We will show numerically that the solution for $r$  switches between the two branches at their point of intersection. That is, at the point in time where,
\begin{equation*}
q^2 + 2s^2\zeta(s) = 0.
\end{equation*}
As we illustrate numerically, this maintains the time-differentiability of $r(t)$.

Notice that $\dot{\nu} > 0$ for all time in a physically realistic solution with $s > 0$. Moreover, from \cref{eqn:HKF}, we can compute
\begin{equation*}
\nu = \frac{-q \pm \sqrt{q^2 - 2s^2\zeta(s)}}{2s^2}.
\end{equation*}
Since $\nu$ varies inversely with $s^2$ and is an increasing function, it follows that when $s$ is small at the boundaries, we may observe stiffness in the ODE, even though solutions exist. That is, the Lagrange multiplier enforcing the variance dynamics is easier to violate numerically. We will observe this in the following examples.

The solutions for $r$ and $\nu$ also imply an existence condition similar to the one in \cref{eqn:qcond}. For an unconstrained optimal control to exist, we must have,
\begin{equation*}
q^2 \geq |2s^2\zeta(s)|.
\end{equation*}
Again, this suggests that strong priors can cause difficulty in changing belief functions.

Finally, in both the geodesic control case and the corrector control case, $z$ tracks $\mu$. However, in the corrector control case the $\mu$ tracking is modified by a $-\Lambda r$ term, where the sign of $\Lambda$ will be determined by the goal $\mu_f$ and its relation to the natural change in $\mu(t)$ as a result of the given value $u$.

\subsection{Examples}

\begin{figure*}[htbp]
\centering
\includegraphics[width=0.24\textwidth]{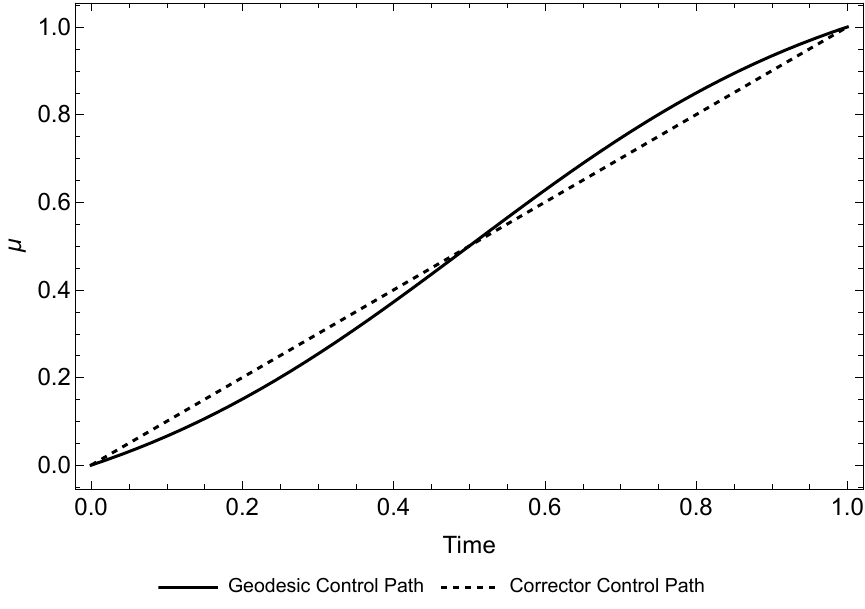}
\includegraphics[width=0.24\textwidth]{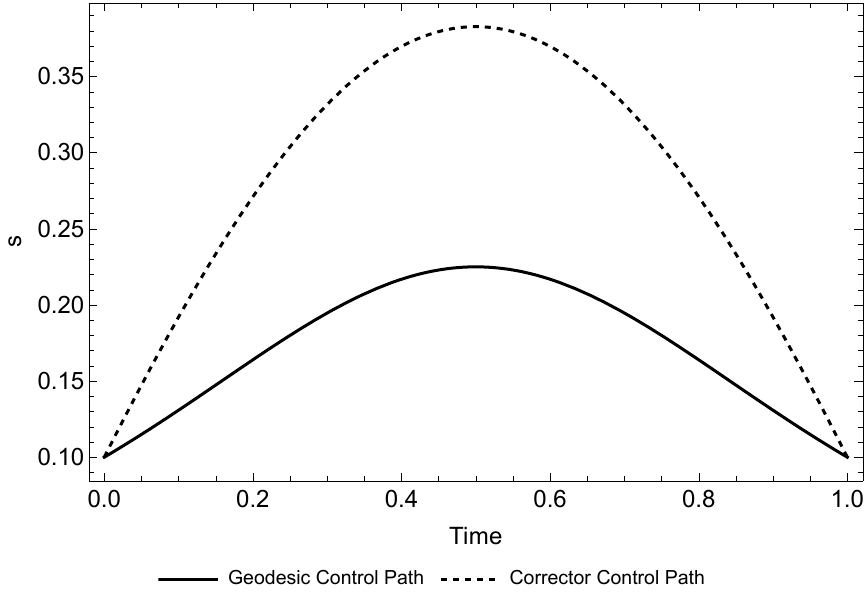}
\includegraphics[width=0.24\textwidth]{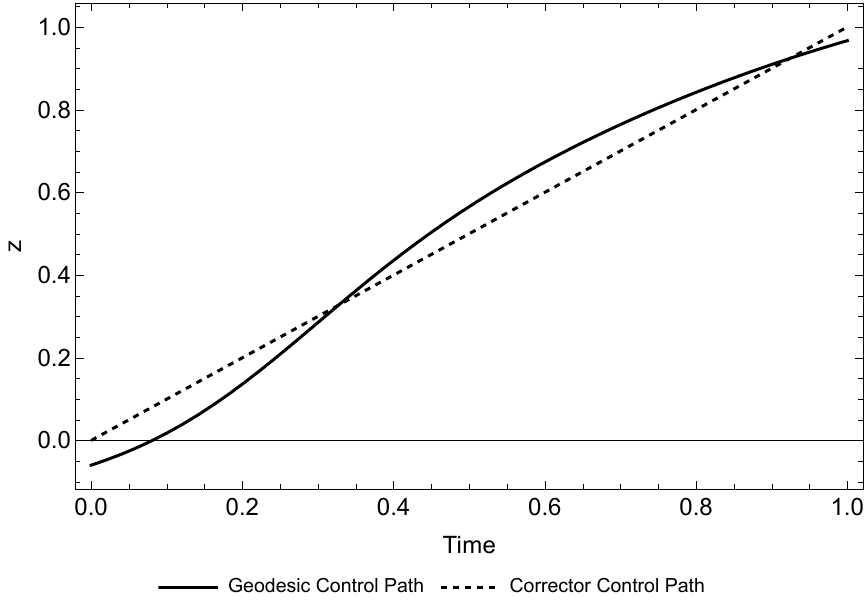}
\includegraphics[width=0.24\textwidth]{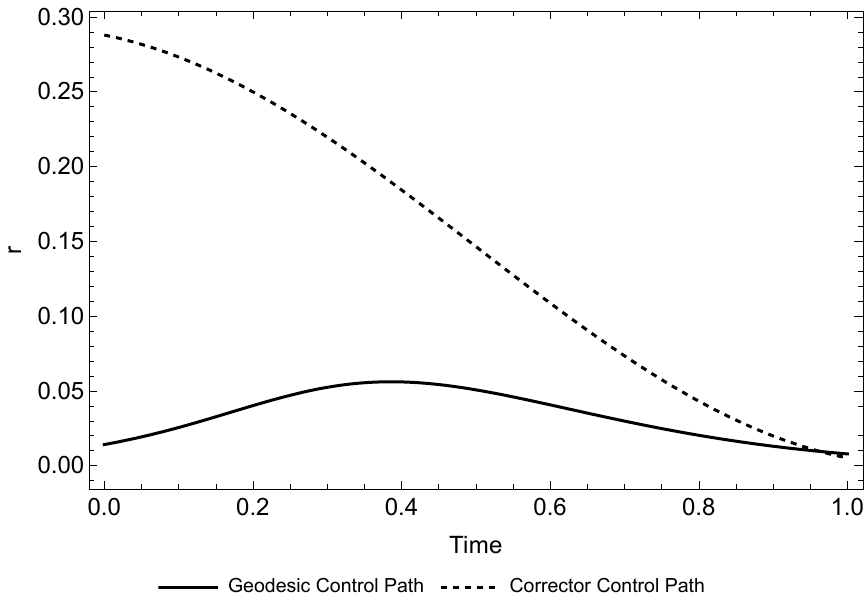}\\
\includegraphics[width=0.24\textwidth]{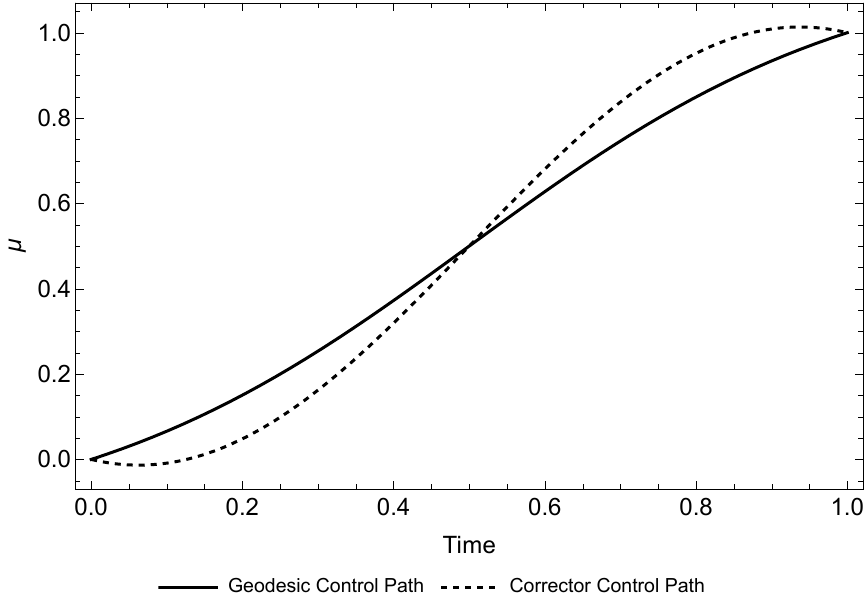}
\includegraphics[width=0.24\textwidth]{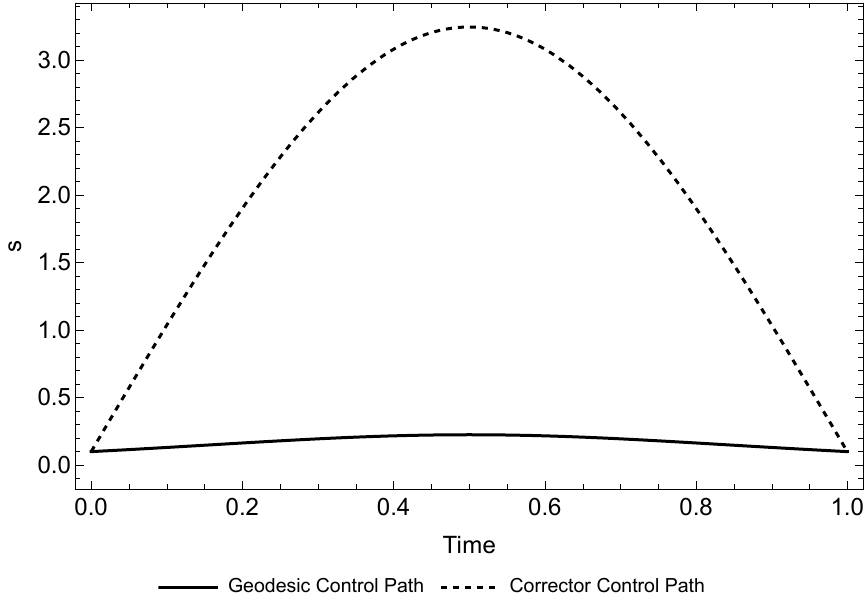}
\includegraphics[width=0.24\textwidth]{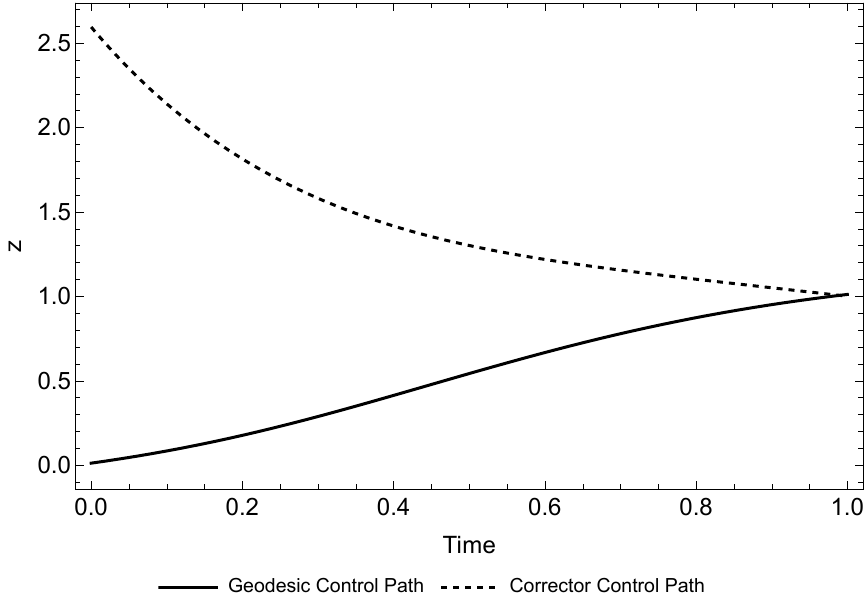}
\includegraphics[width=0.24\textwidth]{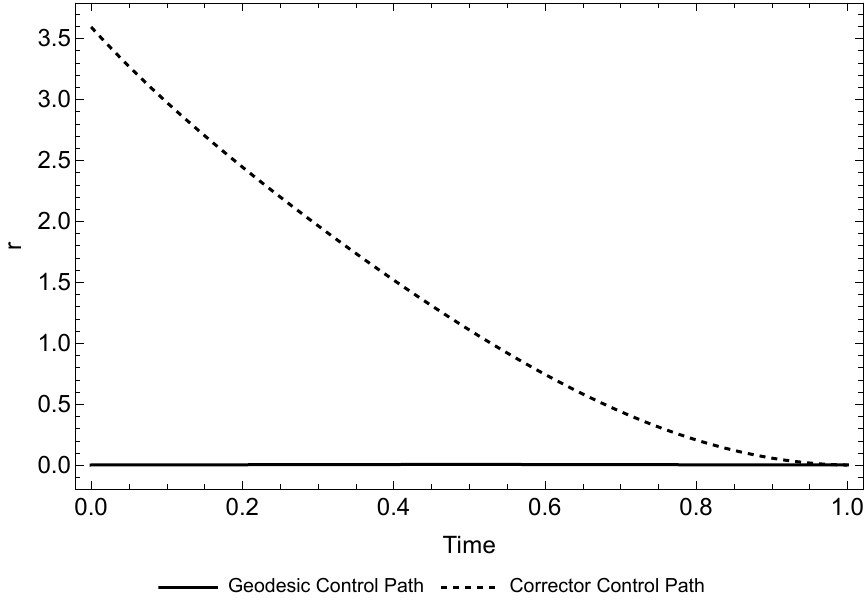}\\
\caption{Example solutions with $\mu(0) = 0$, $\mu(1) = 1$, and $s(0) = s(1) = 0.1$. (Top) The prior information is $u = 1$ and $q = 1$. (Bottom) The prior information is $u = -0.5$ and $q = 9.5$.}
\label{fig:KalmanExample}
\end{figure*}

We compare the geodesic control and the corrector control solutions numerically on two example problems. In both cases we assume $\mu(0) = 0$, $\mu(1) = 1$, and $s(0) = s(1) = 0.1$. That is, the objective is to persuade an observer that $\mu(1) = 1$ and $s(1) = 0.1$, given the initial conditions and the priors, which vary in the two examples.  Solution curves are shown in \cref{fig:KalmanExample}.

In \cref{fig:KalmanExample} (top) we assume the prior information is $u = 1$ and $q = 1$. That is, the assumed motion dynamics are consistent with the desired end state. We see that both the geodesic controller and the corrector controller both successfully drive the system to the desired end state. However, the corrector controller does so by introducing significantly more variance in to the system. In exchange, the corrector controller trajectory for both $\mu$ and $z$ matches the agent's prior assumptions on the evolution of \(\mu\).

\begin{figure}[htbp]
\centering
\includegraphics[width=0.8\columnwidth]{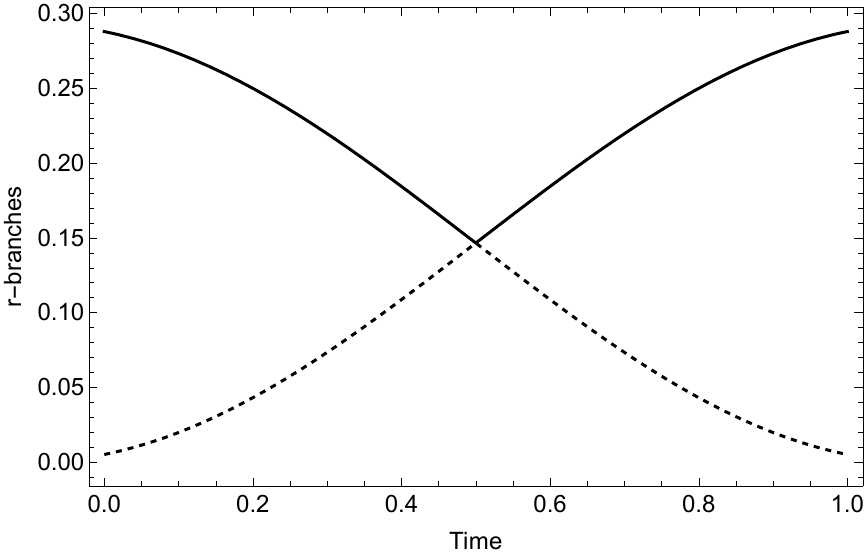}\\
\includegraphics[width=0.8\columnwidth]{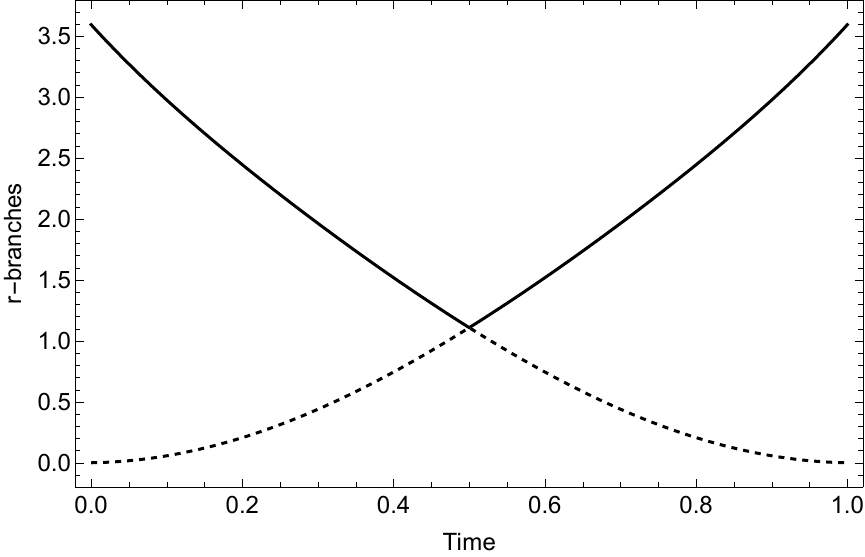}
\caption{Branches of the closed loop control solution for $r$ to be compared to the corresponding rows in \cref{fig:KalmanExample}. The two branches are shown in solid and dashed lines respectively. Note that both exhibit a cusp at $t = 0.5$ where the switch occurs in the controller.}
\label{fig:Branches}
\end{figure}

The two branches of the closed loop solution for $r$ are shown in \cref{fig:Branches} (top). When compared to \cref{fig:KalmanExample} (top, right) it is clear that the solution for $r$ is switching between the two branches at $t = 0.5$.

\begin{figure}[htbp]
\centering
\includegraphics[width=0.8\columnwidth]{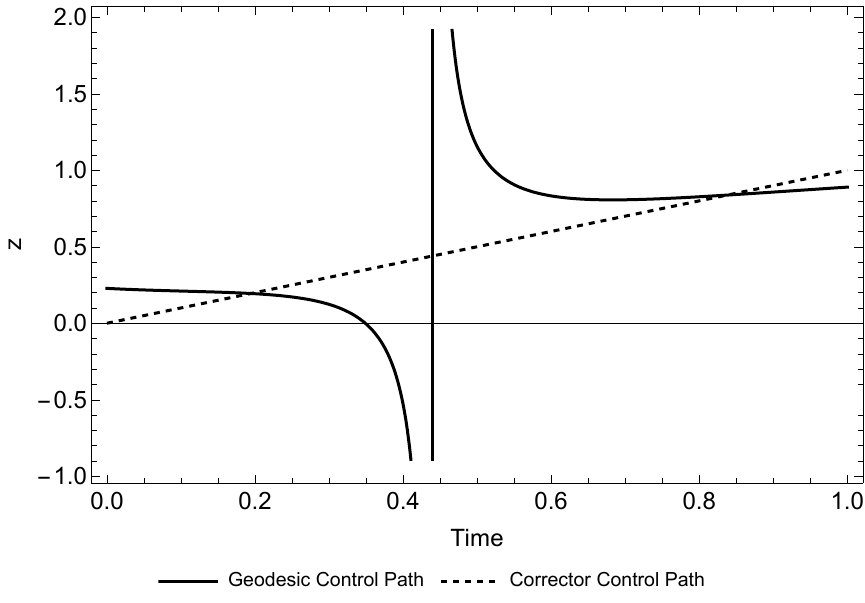}\\
\includegraphics[width=0.8\columnwidth]{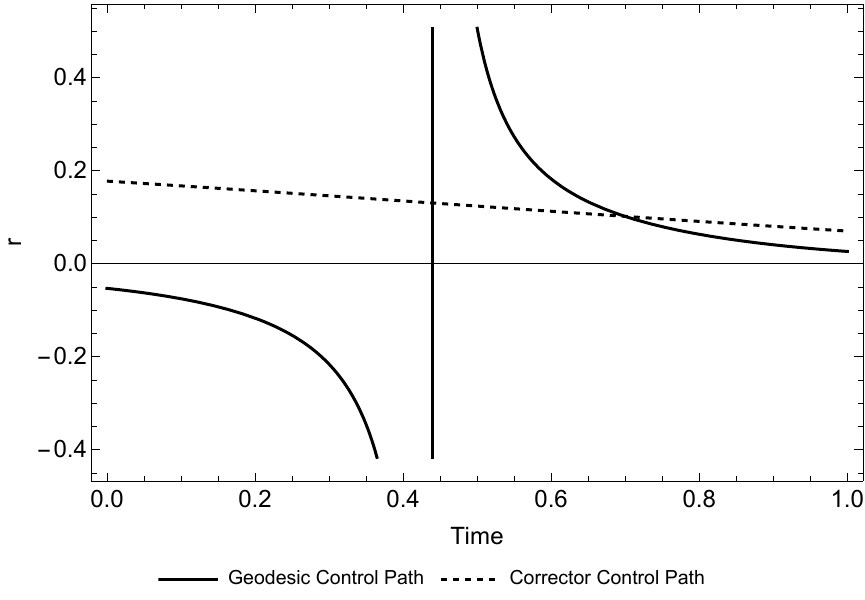}\\
\caption{The solutions for $z$ (top) and $r$ (bottom) in the case when a well-behaved geodesic controller does not exist, but the corrector controller exists.}
\label{fig:rzbad}
\end{figure}

Interestingly, there are parameter choices for which the geodesic control does not exist while the corrector controller does. \cref{eqn:qcond} shows that $q_\text{min} \approx 0.3334$ for the geodesic controller to exist. Setting $q = 0.1$, produces a poorly behaved geodesic controller in which $z$ exhibits blow up in finite time, while $r$ becomes negative. This is shown in \cref{fig:rzbad}.
We contrast this behavior in the next example where parameter classes exist for which the geodesic controller exists but the corrector controller is stiff (and hence a solution exists, but is hard to identify numerically).

In \cref{fig:KalmanExample} (bottom) we assume the prior information is $u = -0.5$ and $q = 9.5$. That is, the assumed dynamics are inconsistent with the desired end state. The prior $q$ is set large to ensure a control solution exists. Numerical methods show that a minimum value for $q$ is $q_\text{min} \approx 8.8343$ to ensure that the corrector controller dynamics are not stiff. If the prior is any stronger, then a numerical solution cannot readily be determined by simple methods.  Compare this with the geodesic controller, where \cref{eqn:qcond} shows that $q_\text{min} \approx 0.3334$ for the geodesic controller to exist. Interestingly, simply adjusting the boundary conditions for $s$ (to be larger) removes this stiffness with all other parameters being held constant (this is not shown).


In general, the geodesic controller produces the same parameter path no matter the a priori belief for the velocity.  Conversely, the corrector controller is sensitive to the agent's priors, and produces the most obvious set of inputs when the desired motion matches said priors. In the case of the inconsistent prior, the corrector controller produces a path that starts with high bias, high variance measurements in order to build uncertainty then transitions to low variance measurements over time. In contrast one might argue that the geodesic controller appears to brute force the system by feeding in low variance measurements that drive the parameters along the geodesic.

\section{Persuasion in a Boltzmann (Softmax) Classifier}\label{sec:Boltzmann}
Boltzmann distributions and Gibbs measures have been studied in the context of information geometry by several authors in relation to thermodynamics \cite{SSBC12,SC12,C07,FC09,D20} as well as neural networks \cite{A97}. In this section, we consider the information geometry of the Boltzmann distribution when treated as a softmax classifier, which measures evidence and provides a probability associated to a specific world state or decision.

Given parameter values $x_i$ and a fixed parameter $\beta$, the discrete Boltzmann distribution is,
\begin{equation}
p_i = \frac{\eexp{\beta x_i}}{\sum_j \eexp{\beta x_j}}.
\label{eqn:Boltzmann1}
\end{equation}
Here $\beta = (kT)^{-1}$ represents the ``inverse temperature'' of the system in the statistical physics sense. This distribution is used as the basis for softmax classifiers \cite{bishop2006} and also appears in $n$-ary options markets, commonly used in prediction markets \cite{GG23}.

We think of $x_i$ as a measure of the evidence supporting conclusion $i$. In this case, controlling the value of $x_i$ can be thought of as providing evidence (information in  bits or nats) persuading/disuaiding an agent that a certain world state is true. From the perspective of misinformation, this may cause a misclassification and thus is related to (but distinct from) problems in adversarial learning (see, e.g., \cite{MXK20}) when the $x_i$ are generated as the penultimate layer of a neural network.

Applying the quotient rule to \cref{eqn:Boltzmann1} yields,
\begin{equation*}
\dot{p}_i = \sum_{j} \beta p_i p_j\left(\dot{x}_i - \dot{x}_j\right).
\end{equation*}
Thus, the dynamics of the distribution depend on the dynamics of the evidence itself. We now assume that the predictor-corrector dynamic is exhibited in the evidence collection process so that,
\begin{equation*}
\begin{bmatrix}
\dot{x}_1\\
\vdots\\
\dot{x}_n
\end{bmatrix} =
\underbrace{\begin{bmatrix}
\phi_1(\mathbf{x})\\
\vdots\\
\phi_n(\mathbf{x})
\end{bmatrix}}_\phi  + \underbrace{\begin{bmatrix}\chi_1(\mathbf{x},\bm{\eta})\\\vdots\\ \chi_n(\mathbf{x},\bm{\eta})\end{bmatrix}}_\chi.
\end{equation*}
This subsumes (a variation of) the leaky integrator version \cite{VLP21} of the diffusion decision model (DDM) \cite{RM08} from mathematical neuroscience in which we have
\begin{equation*}
\dot{x}_i = a_i + \alpha_i x_i + \eta_i.
\end{equation*}
When $\alpha_i < 0$, evidence leaks away (as a process of information aging). If $\alpha_i > 0$, this models confirmation bias \cite{N98} in evidence collection. We note that the DDM model is usually a stochastic differential equation \cite{VLP21,RM08} and defer analysis of this more complex case to future work.

Using Hofbauer's trick \cite{H96}, let
\begin{equation*}
z_i = \frac{p_i}{p_n}\quad \text{and} \qquad q_i = \log(z_i) \iff z_i = \eexp{q_i}.
\end{equation*}
Since $p_1 + \cdots + p_n = 1$, we can write:
\begin{equation}
p_n = \frac{1}{1 + \sum_{i=1}^{n-1} \eexp{q_i}}.
\label{eqn:pn}
\end{equation}
Using the quotient rule we see,
\begin{align*}
\dot{z}_i &= \beta z_i\left(\dot{x}_i - \dot{x}_n\right), \\
\dot{q}_i &= \beta\left(\dot{x}_i - \dot{x}_n\right).
\end{align*}
The Fisher pseudo-metric in terms of $\mathbf{p} = \langle{p_1,\dots,p_n}\rangle$ is,
\begin{equation}
\label{eqn:boltzfr}
\mathbf{g}_{ij} = \begin{cases}
\sum_{j \neq i} \beta^2p_ip_j & \text{if $i=j$}\\
-\beta^2p_ip_j & \text{if $i \neq j$}.
\end{cases}
\end{equation}
The differential form can be computed explicitly, with the Einstein summation convention in the center expression, as
\begin{equation}
\label{eqn:boltzfr2}
ds^2 = \mathbf{g}_{ij}\,dx^i\,dx^j = \sum_{i < j}\beta^2 p_ip_j\left(dx^i - dx^j\right)^2.
\end{equation}
To convert to coordinates in $\mathbf{q}$, note that
\begin{align*}
p_ip_j &= p_n^2 \frac{p_ip_j}{p_n^2} = \frac{\eexp{q_i + q_j}}{\left(1 + \sum_{i=1}^{n-1} \eexp{q_i}\right)^2} \\
p_ip_n &= p_n^2 \frac{p_i}{p_n} = \frac{\eexp{q_i}}{\left(1 + \sum_{i=1}^{n-1} \eexp{q_i}\right)^2}.
\end{align*}
Finally, if $j \neq n$ we have,
\begin{equation*}
(\dot{x}_i - \dot{x}_j) = \frac{\dot{q}_i - \dot{q}_j}{\beta},
\end{equation*}
which gives the Lagrangian for the geodesic control action $A$ as,
\begin{multline}
\mathcal{L}(\mathbf{q},\dot{\mathbf{q}}) = \frac{1}{2\left(1+\sum_{k=1}^{n-1}\eexp{q_k}\right)^2}\cdot \\
\left(\sum_{i < j < n} \eexp{q_i + q_j}(\dot{q}_i - \dot{q}_j)^2 + \sum_{j=1}^{n-1}\eexp{q_j}\dot{q}_j^2\right).
\end{multline}
It is straightforward to construct the action for the corrector control action as,
\begin{multline}
\mathcal{L}_\chi(\mathbf{q},\bm{\chi}) = \frac{1}{2\left(1+\sum_{k=1}^{n-1}\eexp{q_k}\right)^2}\cdot \\
\left(\sum_{i < j < n} \eexp{q_i + q_j}(\chi_i - \chi_j)^2 + \sum_{j=1}^{n-1}\eexp{q_j}\chi_j^2\right),
\end{multline}
by simply replacing the dynamics $\dot{q}_i$ with its corrector component. Without loss of generality, let $\beta = 1$; at this point, it has cancelled out in the Lagrangian and it is simply an extra parameter to carry forward.

\subsection{Predictorless Control Problem}

In the special case when $\dot{x}_i = \eta_i$ (i.e., the predictor is zero), the geodesic and corrector control problems are identical. That is $A = A_\chi$ and we can use classical techniques from the calculus of variations to find the optimal control.

Deriving the Euler-Lagrange equations for an arbitrary initial and final condition with an arbitrary number of states is cumbersome. However, we can derive an exact solution in the case when we begin from the maximum entropy state,
\begin{equation*}
q_1(t_0) = q_2(t_0) = \cdots = q_{n-1}(t_0) = 0
\end{equation*}
and wish to perturb \textit{only} $q_1$ (i.e., we change only $p_1$ while all other states are ignored). In this case, we find the optimal path to the state
\begin{equation*}
q_1(t_f) = a \quad q_2(t_f) = \cdots = q_{n-1}(t_f) = 0.
\end{equation*}
From this we see that $\dot{q}_j = 0$ for $j \neq 1$ and the Lagrangian simplifies to
\begin{equation*}
\mathcal{L}(\mathbf{q},\dot{\mathbf{q}}) = \frac{(n-1)\eexp{q_1}\dot{q}_1^2}{2\left[(n-1) + \eexp{q_1}\right]^2}.
\end{equation*}
Computing the Euler-Lagrange equation and solving for $\ddot{q}_1$ yields
\begin{equation*}
\ddot{q}_1 = \frac{\left[\eexp{q_1} +1-n\right]\dot{q}_1^2}{2\left[(n-1) + \eexp{q_1}\right]},
\end{equation*}
which can be phrased as the system of first order differential equations using the control variable $\eta_1$,
\begin{equation*}
\left\{
\begin{aligned}
&\dot{\eta} = \frac{\left[\eexp{q} +1-n\right]\dot{q}^2}{2\left[(n-1) + \eexp{q}\right]}\\
&\dot{q} = \eta\\
&q(0) = 0, q(T) = a,
\end{aligned}
\right.
\end{equation*}
where we remove the subscripts for readability. Because of the simplicity of the Boltzmann distribution, the units of $\eta$ are naturally in nats (or bits) per time unit.

This system of differential equations has a closed form solution:
\begin{align}
&q(t) = \log \left((n-1) \tan ^2\left(\frac{1}{2} \sqrt{n-1} (c_1 t+c_2)\right)\right)\\
&
\eta(t) = c_1 \sqrt{(n-1) \tan ^2\left(\frac{1}{2} \sqrt{n-1} (c_1 t+c_2)\right)} \cdot \nonumber\\
&\hspace*{6em} \csc
   ^2\left(\frac{1}{2} \sqrt{n-1} (c_1 t+c_2)\right).\label{eqn:uboltz}
\end{align}
If we assume $t_f = 1$, the constants are given by,
\begin{align*}
&c_1 = \frac{\cos ^{-1}(1-2 a)-\cos ^{-1}\left(\frac{n-2}{n}\right)}{\sqrt{n-1}},\\
&c_2 = \frac{\cos ^{-1}\left(\frac{n-2}{n}\right)}{\sqrt{n-1}}.
\end{align*}
Using the fact that $q_2 = \cdots q_{n-1} = 0$ and \cref{eqn:pn}, we can compute
\begin{align}
p_1(t) &= \sin ^2\left(\frac{1}{2} \sqrt{n-1} (c_1 t+c_2)\right), \\
p_2(t) &= \cdots = p_n(t) = \frac{\cos ^2\left(\frac{1}{2} \sqrt{n-1} (c_1 t+c_2)\right)}{n-1}.
\end{align}

Further, if we assume that $p_1(t_f) > \tfrac{1}{n}$ (i.e., $q(t_f) > 0$), then $\eta(t) > 0$ and we use the positive branch of the square root in \cref{eqn:uboltz} to find the simpler expression,
\begin{equation*}
\eta(t) = 2 c_1 \sqrt{n-1} \csc \left(\sqrt{n-1} (c_1 t+c_2)\right).
\end{equation*}
If the opposite is true, then we simply use the negative branch and $\eta(t)$ is the negative of this quantity.

\begin{figure}[htbp]
\centering
\includegraphics[width=0.9\columnwidth]{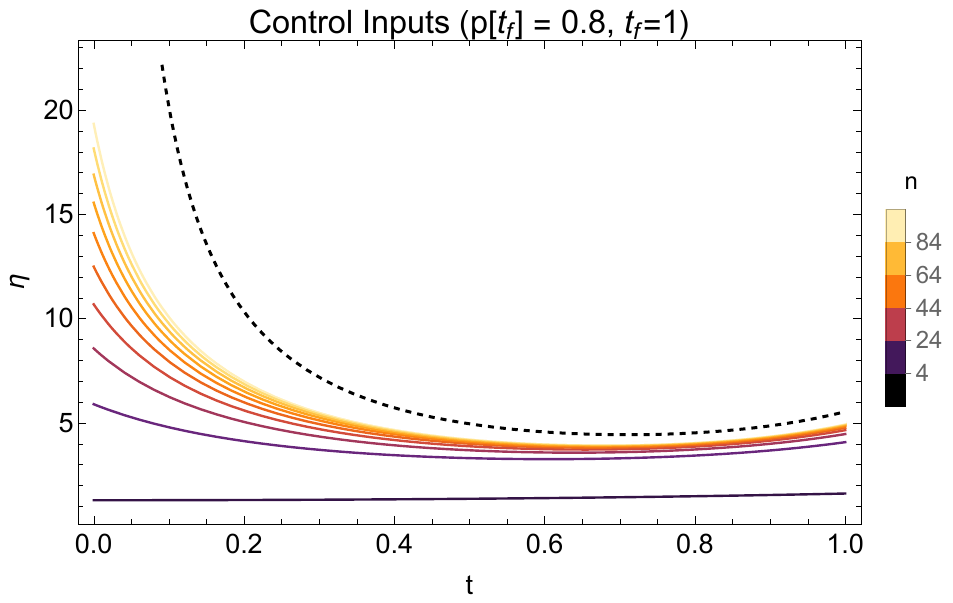}\\
\includegraphics[width=0.9\columnwidth]{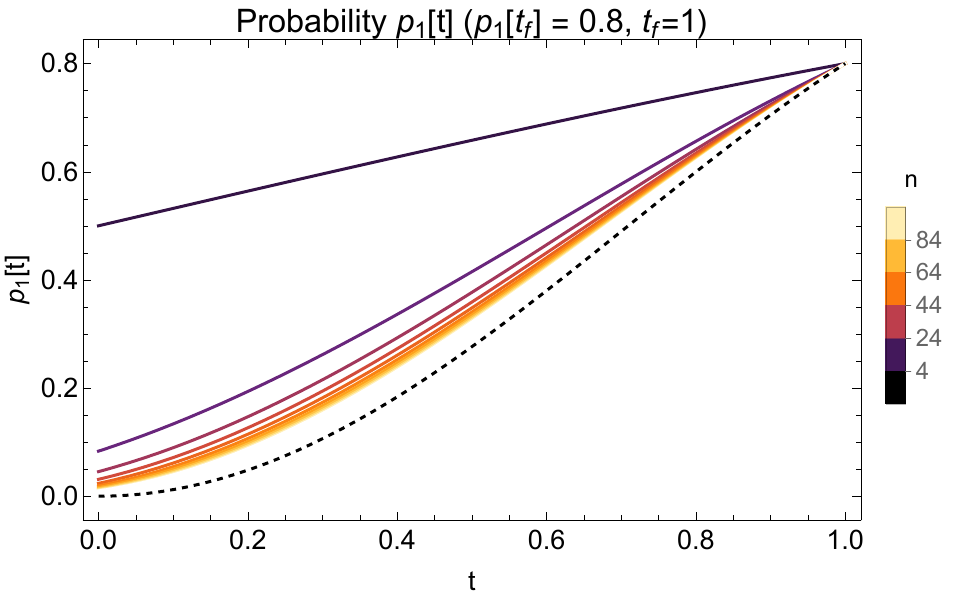}
\caption{(Top) The control curves for various $n$ ($2 \leq n \leq 102$). The black dashed control curve is the asymptotic control when $n \to \infty$. (Bottom) The trajectories of the probability of belief in State 1 given control inputs for various $n$ (as specified). The black dashed curve is the asymptotic behavior of $p_1(t)$ as $n\to\infty$. In both figures $t_0 = 0$, $t_f = 1$, and $p_1(t_f)=0.8$.}
\label{fig:Curves}
\end{figure}

In the edge case when $n = 2$, these results simplify substantially to,
\begin{align*}
p_1(t) &= \sin ^2\left(\frac{1}{2} (c_1 t+c_2)\right), \\
p_2(t) &= \cos ^2\left(\frac{1}{2} (c_1 t+c_2)\right), \\
\eta(t) &= 2 c_1 \csc (c_1 t+c_2).
\end{align*}
We can likewise compute the explicit form of the solution when $n \to \infty$, representing the case of many possible belief states. Then we have,
\begin{align*}
&p_1(t) = \sin ^2\left(\frac{1}{2} t \cos ^{-1}(1-2 a)\right)\\
&\eta(t) = 2 \cos ^{-1}(1-2 a) \csc \left(t \cos ^{-1}(1-2 a)\right).
\end{align*}
The control curves and probability paths for a range of $n$ when $p_1(t_f)=0.8$ are illustrated in \cref{fig:Curves}.

Note that these plots do not directly show the path in the parameter space $x_i$.  Instead we see how the resulting probability changes over time and the rate $\eta$ that information must be supplied to change the probability.  First observe that, like previous examples, the rate of change of the probability is fastest when the variance is high  and slows near areas of certainty.  This is consistent with the Fisher-Rao metric and our expectation that regions of certainty will have higher curvature.  On the other hand, the rate at which information must be funneled into the system has its maxima at $t=t_0$ and $t=t_f$.  Indeed the region when the probability is changing the fastest is also the region where the information input is smallest. The nature of the Boltzmann distribution requires increasing energy expenditure to shift distributions towards further certainty.  More broadly, reduced utility for persuasion targeted towards strong priors is an important dynamic.

\subsection{Leaky Integrator Control Problem}

Analyzing the control problem for the Boltzmann distribution when inputs have fully general dynamics requires numerical analysis. However, we can gain insight from the special case when for all $i$, $x_i$ follows the dynamic
\begin{equation*}
\dot{x}_i = \underbrace{a + \alpha x_i}_{\phi_i} + \underbrace{\eta_i}_{\chi_i}.
\end{equation*}
That is, we assume all information dynamics share a common predictor model and only the corrector (control) dynamics vary. For simplicity, we will continue to assume that $\beta = 1$.

From our previous analysis, we see that
\begin{equation*}
\dot{q}_i = \alpha(x_i - x_n) + (\eta_i - \eta_n).
\end{equation*}
Let $\xi_i = x_i - x_n$ and $\zeta_i = \eta_i - \eta_n$. Notice that
\begin{equation*}
\dot{\xi}_i = \dot{x}_i - \dot{x}_n = \dot{q}_i.
\end{equation*}
Therefore $q_i = \xi_i + C_i$ for some constant $C_i$. If we impose a symmetry assumption that all $x_i(0)$ are equal, then initially $q_i(0) = 0 = \xi_i(0)$ for all $i$. It follows at once that $C_i = 0$ and $q_i = \xi_i$ and $q_i$ has dynamics
\begin{equation*}
\dot{q}_i = \alpha q_i + \zeta_i.
\end{equation*}
That is, in this highly symmetric case, $q_i$ inherits the dynamics of $x_i$ almost entirely.

If we again assume that we wish to modify only $q_1$ then for $i \neq 1$, as before, $\eta_i = 0$, which implies $\zeta_i = 0$. Since we assumed that we are starting from a maximum entropy state, we have $q_i(0) = 0$ and, therefore, $q_i(t) = 0$ for all time, since $\dot{q}_i = 0$. For $i = 1$, we have $\zeta_1 = \eta_1$ and the resulting simplified Lagrangians for the geodesic case and corrector case in terms of $\eta_1$ are:
\begin{align*}
\mathcal{L}(\mathbf{q},\bm{\zeta}) = \frac{(n-1)\eexp{q_1}(\alpha q_1 + \eta_1)^2}{2\left[(n-1) + \eexp{q_1}\right]^2}\\
\mathcal{L}_\chi(\mathbf{q},\bm{\zeta}) = \frac{(n-1)\eexp{q_1}\eta_1^2}{2\left[(n-1) + \eexp{q_1}\right]^2}.
\end{align*}
The dynamics of $q_1$ are then,
\begin{equation*}
\dot{q}_1 = \alpha q_1 + \zeta_1 = \alpha q_1 + \eta_1.
\end{equation*}

For simplicity, we will discard the subscripts. The Hamiltonians for the two control systems are almost identical,
\begin{align*}
&H_\text{geodesic} = \frac{(n-1)\eexp{q}(\alpha q + \eta)^2}{2\left[(n-1) + \eexp{q}\right]^2} + \lambda \left(\alpha q + \eta \right)\\
&H_\text{corrector} = \frac{(n-1)\eexp{q}\eta^2}{2\left[(n-1) + \eexp{q}\right]^2} + \lambda \left(\alpha q + \eta \right)
\end{align*}
The resulting dynamics are similar with,
\begin{equation*}
\text{Geodesic}\left\{
\begin{aligned}
\dot{q} &= -\frac{\lambda  e^{-q} \left(n-1+e^q\right)^2}{n-1}\\
\dot{\lambda} &= \frac{\lambda^2\left(n-1+\eexp{q}\right)}{2(n-1)}\left(1-(n-1)\eexp{-q}\right)
\end{aligned}\right.
\end{equation*}
and
\begin{equation*}
\text{Corrector}\left\{
\begin{aligned}
\dot{q} &= \alpha q -\frac{\lambda  e^{-q} \left(n-1+e^q\right)^2}{n-1}\\
\dot{\lambda} &= -\alpha\lambda + \frac{\lambda^2\left(n-1+\eexp{q}\right)}{2(n-1)}\left(1-(n-1)\eexp{-q}\right),
\end{aligned}\right.
\end{equation*}
where the controls are given by,
\begin{align*}
&\eta_\text{geodesic} = -\alpha q -\frac{\lambda  e^{-q} \left(n-1+e^q\right)^2}{n-1}\\
&\eta_\text{corrector} = -\frac{\lambda  e^{-q} \left(n-1+e^q\right)^2}{n-1}\\
\end{align*}

Notice that $\alpha$ does not appear in the geodesic dynamics. As expected, the geodesic path is not affected by information leakage or bias. This is accounted for by the controller itself. On the other hand, the behavior of the corrector dynamics does include $\alpha$ and consequently the corrector controller will modify the path through probability space to take advantage of (or to compensate for) leakage or bias, even though the leakage/bias term does not appear explicitly in the control law.

The geodesic dynamics do have a (large) closed form solution in elementary functions, but it is not useful as a point of comparison since the corrector controller does not appear to have a convenient closed form solution, even though the system is integrable. A more useful comparison is by numerical example, where we show that the information injected into the system is always less in the corrector controller case (as expected).

\begin{figure*}
\centering
\includegraphics[width=0.3\textwidth]{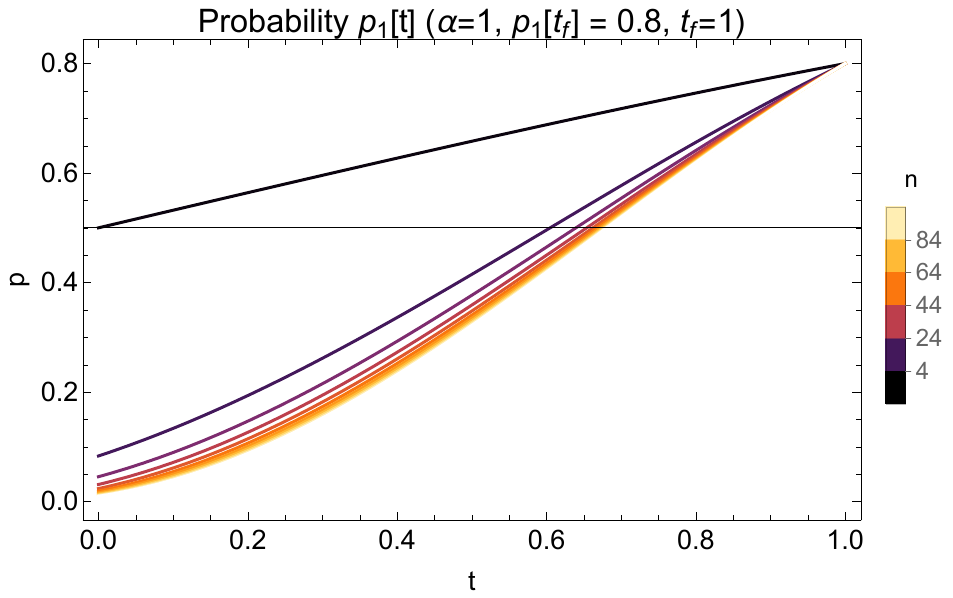}
\includegraphics[width=0.3\textwidth]{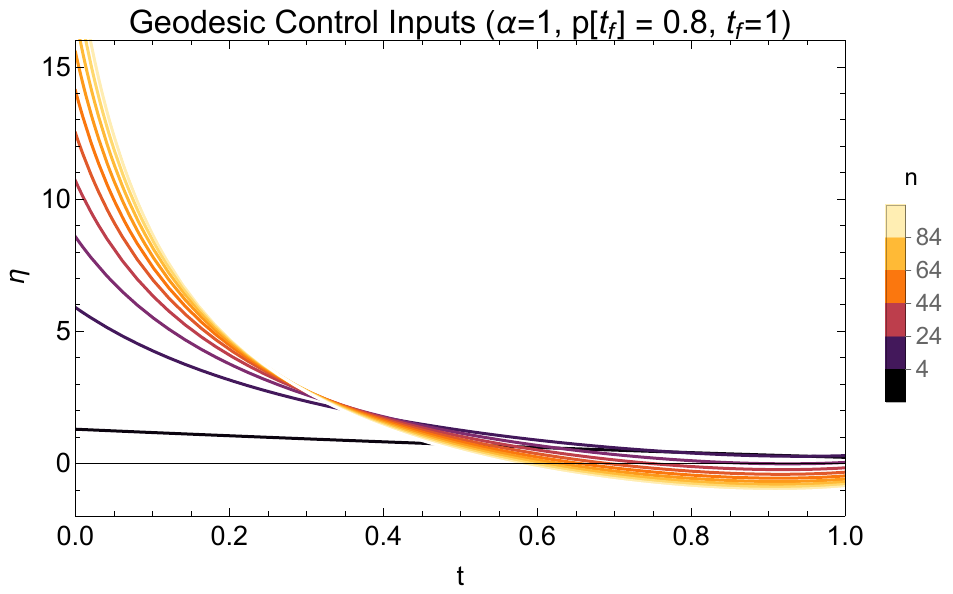}
\includegraphics[width=0.3\textwidth]{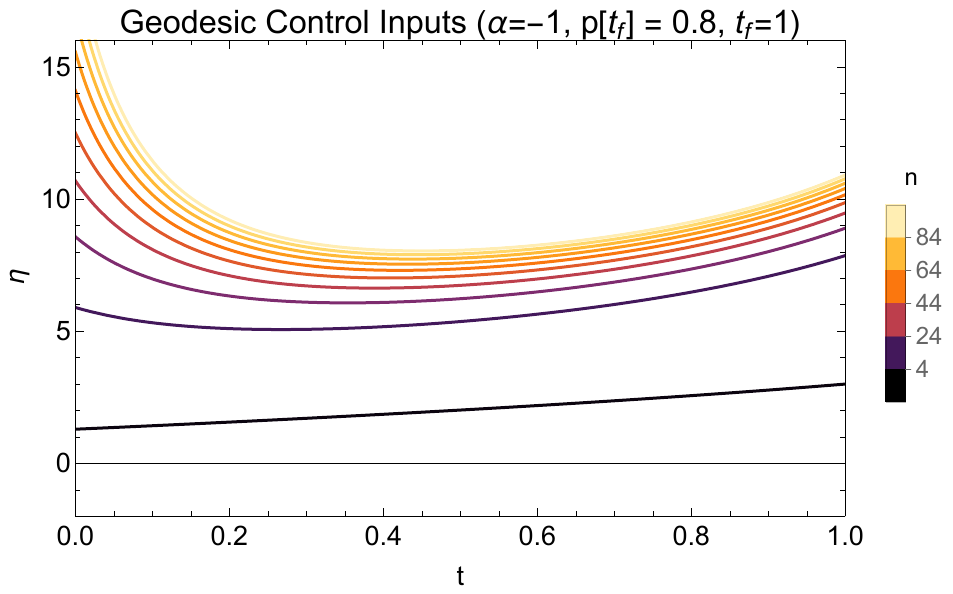}\\
\includegraphics[width=0.3\textwidth]{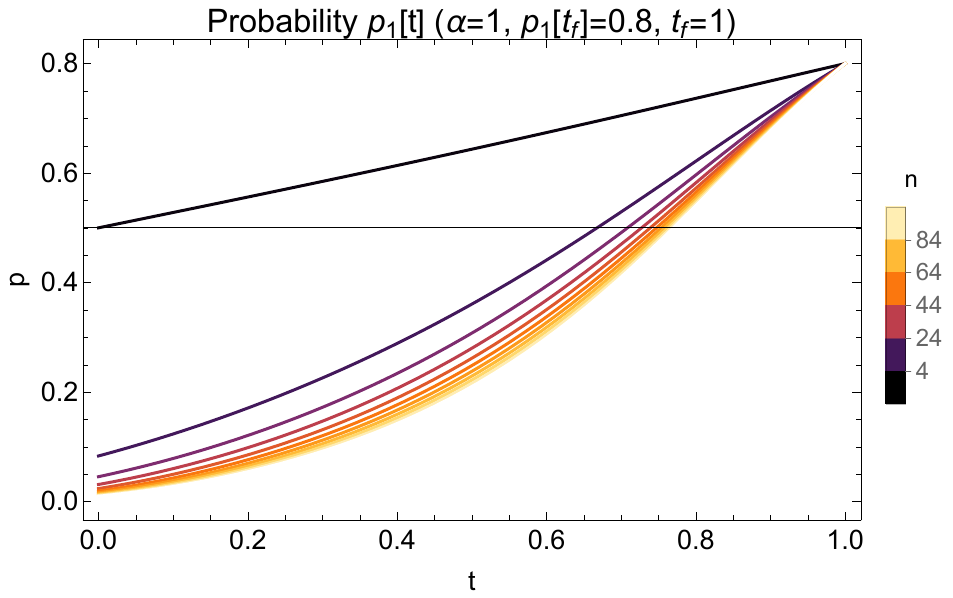}
\includegraphics[width=0.3\textwidth]{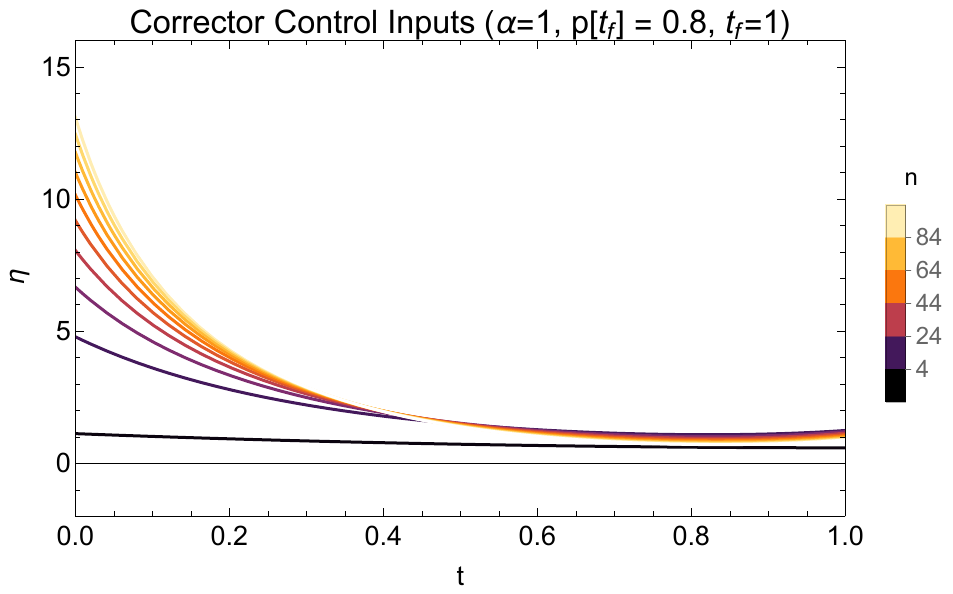}
\includegraphics[width=0.3\textwidth]{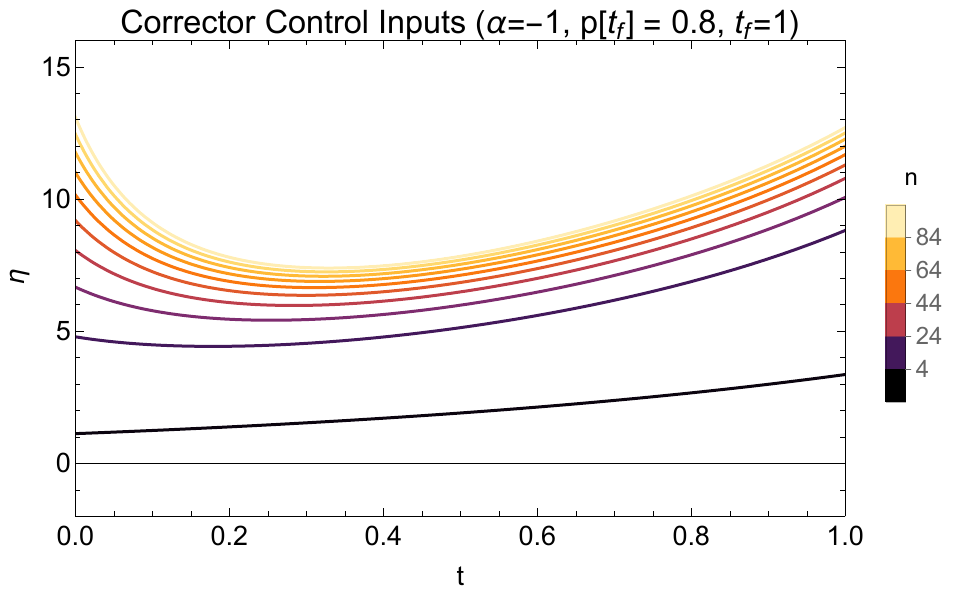}
\caption{(Left) Controlled probability paths $p_1$ for (top) geodesic controller, (bottom) corrector controller. (Center) Controller inputs $\eta$ for (top) geodesic controller, (bottom) corrector controller, when $\alpha = 1$, modeling bias. (Right) Controller inputs $\eta$ for (top) geodesic controller, (bottom) corrector controller, when $\alpha = -1$, modeling information leakage.
}
\label{fig:BoltzmannWithBias}
\end{figure*}

\cref{fig:BoltzmannWithBias} shows the probability path for $p_1(t)$ and the control signals for the geodesic and corrector controllers (both in nats per time unit) for the cases when $\alpha = 1$ and $\alpha = -1$.
It is worth noting that only one probability path is shown because both controllers use a single path.  This is expected for the geodesic controller, but appears to be a quirk of the corrector controller, whose solutions are invariant to the sign of $\alpha$. In all cases, the information input per unit time into the system  is less for the corrector controller than it is for the geodesic controller.

Interestingly, in the case when $\alpha = 1$, the information input in the geodesic controller actually becomes negative (which is allowed in the DDM) to compensate for the biasing affect and to keep the probability following the geodesic. This does not happen in the corrector controller case, which uses the bias to its advantage. In a real-world scenario, negative information would correspond to some kind of counter-factual evidence supporting a specific world state (or information supporting one of the other $n-1$ world states). In the case when $\alpha = -1$, both controllers decrease and then increase their information rates to compensate for the information leakage.

\section{A Rudimentary Fusion Model}\label{sec:Fusion}
The building blocks described above can be combined to produce more complicated models using the underlying information geometry.  An important fact is that given two parameterized probability distributions $p_1(\mathbf{x}|\bm{\theta})$ and $p_2(\mathbf{x}|\bm{\theta})$ with associated Fisher-Rao metrics $\mathbf{g}^1$ and $\mathbf{g}^2$, respectively, then the Fisher-Rao metric for the product distribution $p_1(\mathbf{x}_1|\bm{\theta})p_2(\mathbf{x}_2|\bm{\theta})$ is given by
\begin{displaymath}
\mathbf{g}(\bm{\theta}) = \mathbf{g}^1(\bm{\theta}) + \mathbf{g}^2(\bm{\theta}).
\end{displaymath}
This follows from direct computation and the fact that
\begin{displaymath}
\mathbb{E}_{\bm{\theta}}\left[\frac{\partial p(\mathbf{x}|\bm{\theta})}{\partial \bm{\theta}}\right] = 0.
\end{displaymath}
Restated in terms of actions and Hamiltonians, this implies that for independent joint distributions we can simply add the actions (geodesic or corrector) associated to each distribution.

By way of example, assume there are $N$ independent one-dimensional Kalman filters as in \cref{sec:Kalman} with belief parameters $(\mu_k,s_k)$, observed mean and variance and $(z_k,r_k)$, acting as control variables, and priors $(u_k, q_k)$. Here $k\in\{1,\dots,N\}$. For filter $k$ we have a Hamiltonian $H_k$ given by \cref{eqn:KalmanHamiltonianCorrector}, with the appropriate set of parameters.

An individual may wish to know which (if any) of the Kalman trackers represents a real trajectory (vs. spuriously coherent noise). Such a scenario may occur, e.g., in the visual systems of predators who are attempting to isolate prey trajectories from clutter. (See \cite{XCBL13}, for supporting evidence that Kalman filters are adequate models of visually guided pursuit dynamics.)

A reasonable assumption is that the agent is using the variance of the perceived position as evidence for a Boltzmann distribution indicating which of the Kalman filters represents the more likely (real) trajectory.  Specifically suppose the probability that filter $k$ is the true track is given by \cref{eqn:Boltzmann1} where the evidence for filter $k$ is the negative entropy associated to the observed position.  While the entropy of a normal distribution with variance $s$ is given by $\frac{1}{2}\log(2\pi e s)$, we will disregard the unimportant constant factors and define
\begin{equation}
x_k = -\log(s_k).
\label{eqn:Fusionxk}
\end{equation}
In this model, more variance leads to higher entropy implying a smaller probability of associating a signal with a true track.

Using \cref{eqn:sdot}, we can split the time derivative of $x_k$ into predictor and corrector terms as,
\begin{displaymath}
\dot{x}_k = \frac{-\dot{s}_k}{s_k} = \underbrace{\frac{-q_k}{s_k}}_\phi + \underbrace{\frac{s_k}{r_k}}_\chi,
\end{displaymath}
with the control variable being $r_k$ as before.
As we are taking a numerical approach there is no need to use Hofbauer's trick in this case. The Hamiltonian associated to the Boltzmann distribution can be read directly from \cref{eqn:Boltzmann1,eqn:boltzfr2,eqn:Fusionxk} as,
\begin{multline*}
H_0 = \sum_{i<j} \beta^2 p_i p_j \left( \frac{s_i}{r_i} - \frac{s_j}{r_j}\right)^2
 = \\
 \sum_{i<j} \frac{ \beta^2 s_i^{-\beta} s_j^{-\beta}}{\left(\sum_k s_k^{-\beta}\right)^2} \left( \frac{s_i}{ r_i} - \frac{s_j}{r_j}\right)^2.
\end{multline*}
Note this Hamiltonian has no Lagrange multipliers because we have used \cref{eqn:Boltzmann1} to remove $p_i$ from the expression. The total system Hamiltonian is then,
\begin{equation}
\label{eqn:combham}
H = H_0 + \sum_{k} H_k.
\end{equation}
Using \cref{eqn:combham} we can write down a the Euler-Lagrange differential equations, which can be solved numerically and which provide necessary conditions for an optimal control.

\begin{figure*}
\centering
\includegraphics[width=0.3\textwidth]{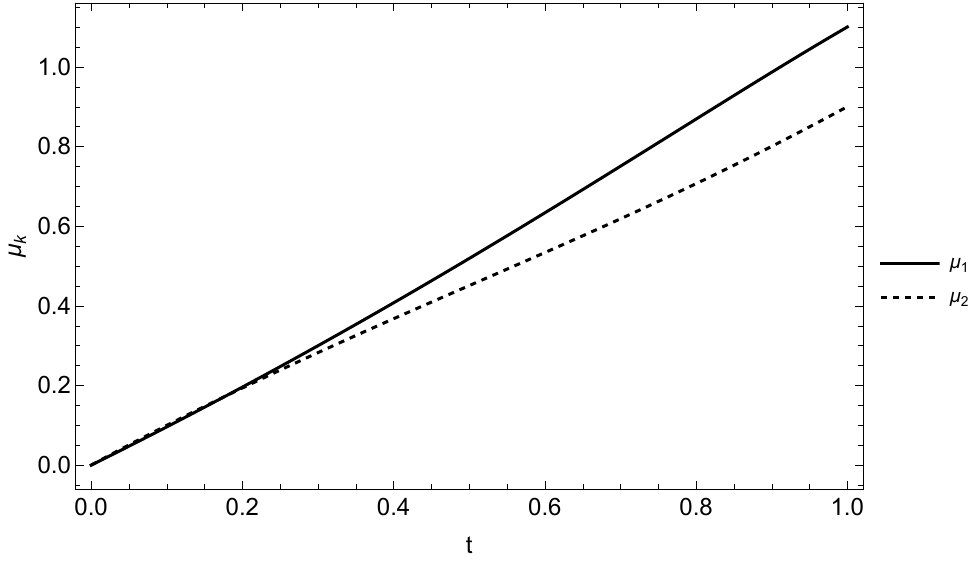}
\includegraphics[width=0.3\textwidth]{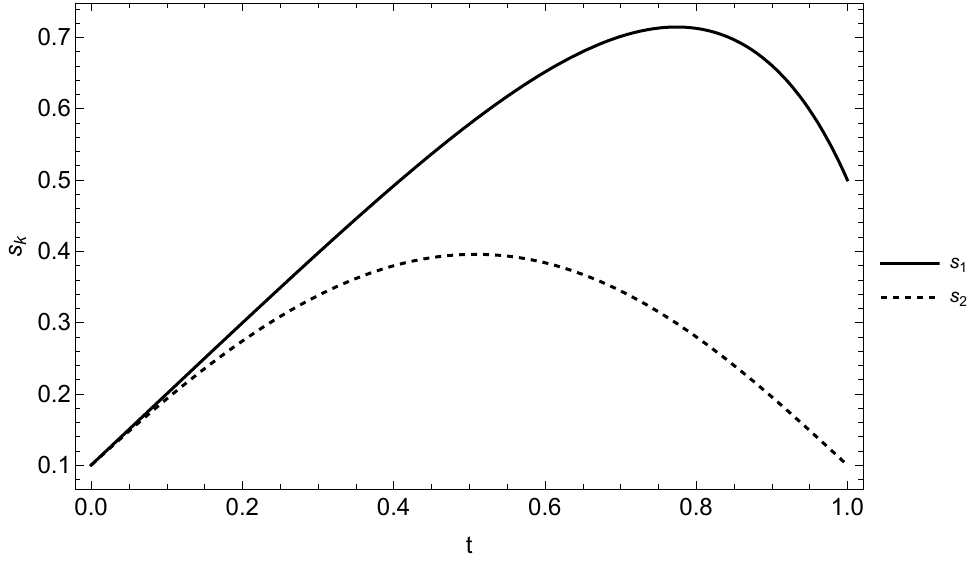}
\includegraphics[width=0.3\textwidth]{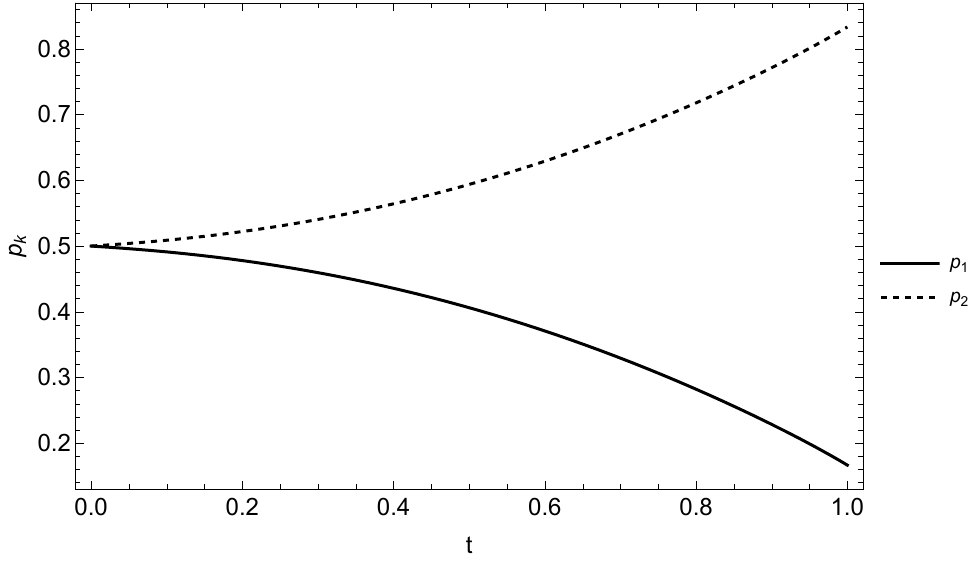}
\caption{(Left) The optimal mean values $\mu_1$ and $\mu_2$. (Center) The optimal variances $s_1$ and $s_2$. (Right) The probability of assuming path $i$ contains a true trajectory.}
\label{fig:MultiKalman}
\end{figure*}

By way of example, consider a scenario in which the objective is to generate inputs to persuade an observer that one of two trajectories is real, subject to prior assumptions on the Kalman filters.  Assume that the agent has prior belief dynamics defined by $\beta = q_k = 1$ and $u_1 = 0.9$ and $u_2 = 1.1$.  Suppose also that we have the initial conditions
\begin{align*}
\mu_1(0) = 0 \quad s_1(0) = 0.1\\
\mu_2(0) = 0 \quad s_2(0) = 0.1.
\end{align*}
Our objective can be accomplished (among other methods) by achieving the final conditions,
\begin{align*}
\mu_1(t_f) = 0.9 \quad s_1(t_f) = 0.5\\
\mu_2(t_f) = 1.1 \quad s_2(t_f) = 0.1,
\end{align*}
which leads to $p_1(t_f) \approx 0.167$ and $p_2(t_f) \approx 0.833$. Thus, the second trajectory has a substantially higher probability of being identified as real. The resulting dynamics are shown in \cref{fig:MultiKalman}.

While the results are largely unsurprising, there are some interesting characteristics. First note that the optimal mean values $\mu_k$ follow linear paths and, as a consequence of the desired final state agreeing with the prior assumptions on $u_k$, the control inputs $z_k$ follow the same paths (not shown).  Also, there is an asymmetry in the $s_1$ variance, which has not been seen before and is most likely being induced by the introduction of the $H_0$ Hamiltonian elements. As before, however, variance increases and then decreases, with the uncertainty adjusting the curvature of the manifold and thus impacting the distance traversed. The probability results are unintuitive in that the rate of change of the probability appears to be accelerating in regions of higher certainty, where the metric will be more costly. However, this may be a function of choosing a small $\beta$, which (in essence) is now acting as a conversion parameter between the cost associated with the Fisher metric for the Boltzmann distribution and the Fisher metrics for the Kalman filters. Further study of these joint metrics is a direction of future research.

\section{Conclusions and Future Directions}\label{sec:Future}
In this paper, we present a mathematical model of optimal persuasion, based on psychological models used in learning theory and in understanding misinformation. Our approach formalizes the required notion of familiarity by phrasing the problem in terms of the manipulation of a belief distribution. Familiarity is then couched in terms of free entropy and the Fisher-Rao metric is both used and modified to characterize a path of `greatest familiarity' (least surprise). In particular, we develop a modified Lagrangian that takes into account natural time-varying dynamics of the belief probability distribution in the absence of intentional control (or manipulation). We compare control signals produced using the Fisher-Rao metric to this modified metric for two classic distributions: the Kalman filter and the Boltzmann distribution, modeling an agent attempting to classify a world state. We then showed how to combine these systems to model a rudimentary information fusion system for a single real object in a cluttered field.

There are several future directions of research that may result from this work. In the first case, a closer integration of the mathematical models with classical models from the neuroscience literature, including those incorporating stochasticity would result in a richer understanding of learning/influence in decision making circumstances. This could include experimental validation of the underlying Fisher-Rao metric as a reasonable (component of) a metric for optimizing persuasion. Other richer models may also consider additional terms in the objective function of the control problem. These terms might incorporate the cost of injecting information into the system (either as a part of the learning process or a malicious misinformation process). Relating this work to (online) adversarial learning in machine learning may also be a fruitful area of research. Additionally, further study of joint distribution models may lead to insights into learning or misinformation in complex decision frameworks. Finally, an empirical study of simple misinformation or learning in the context of the (time-varying) distributions presented here would determine the validity of this model and our underlying assumption that the Fisher-Rao metric is a good representation of neurological ``familiarity'' and thus relevant in learning or influence.

\section*{Acknowledgement}
G. G. and C. G. were supported in part by the Defense Advanced Research Projects Agency through NavSea Task Description DO 21F8366 and Contract HR0011-22-C-0038.

\appendix

\section{Review of Optimal Control}\label{app:OC}
We present key facts from optimal control theory used in this study. Details are available in \cite{Mang66,K04,Friesz10}.

A Bolza type optimal control problem is an optimization problem of the form:
\begin{equation}\left\{
\begin{aligned}
\min\;\; & \Psi(\mathbf{x}(t_f)) + \int_{t_0}^{t_f} f(\mathbf{x}(t), \mathbf{u}(t),t) dt\\
s.t.\;\; & \mathbf{\dot{x}} = \mathbf{g}(\mathbf{x}(t), \mathbf{u}(t),t)\\
&\mathbf{x}(0) = \mathbf{x}_0
\end{aligned}\right.
\label{eqn:Bolza}
\end{equation}
When $\Psi(\mathbf{x}(t_f)) \equiv 0$, this is called a Lagrange type optimal control problem. The vector of variables $\mathbf{x}$ is called the state, while the vector of decision variables $\mathbf{u}$ is called the control. Additional constraints on $\mathbf{u}$, $\mathbf{x}$ or the joint function of $\mathbf{x}$ and $\mathbf{u}$ can be added.

The \textit{Hamiltonian} with adjoint variables (Lagrange multipliers) $\boldsymbol{\lambda}$ for this problem is:
\begin{displaymath}
\mathcal{H}(\mathbf{x},\boldsymbol{\lambda}, u) = f(\mathbf{x}(t), \mathbf{u}(t),t) + \boldsymbol\lambda^T\mathbf{g}(\mathbf{x}(t), \mathbf{u}(t),t)
\end{displaymath}

In what follows, we assume that all $f(\mathbf{x},\mathbf{u},t)$ and $\mathbf{g}(\mathbf{x},\mathbf{u},t)$ are continuous and differentiable in $\mathbf{x}$ and $\mathbf{u}$ and $\Psi(\mathbf{x}(t_f))$ is continuous and differentiable in $x(t_f)$. A proof of the following lemma can be found in almost every book on optimal control (e.g. \cite{Friesz10}).

\begin{lemma}[Necessary Conditions of Optimal Control] If $\mathbf{u}^*$ is a solution to Optimal Control Problem (\ref{eqn:Bolza}), then there is a vector of adjoint variables $\boldsymbol{\lambda}^*$ so that:
\begin{equation}
\mathcal{H}(\mathbf{x}^*(t), \mathbf{u}^*(t), \boldsymbol{\lambda}^*(t)) \leq \mathcal{H}(\mathbf{x}^*(t), \mathbf{u}(t), \boldsymbol{\lambda}^*(t))
\end{equation}
for all $t \in [0, T]$ and for all admissible inputs $\mathbf{u}$, and the following conditions hold:
\begin{enumerate}
\item Pontryagin's Minimim Principle: $\frac{\partial \mathcal{H}}{\partial \mathbf{u}} = \mathbf{0}$ and $\frac{\partial^2\mathcal{H}}{\partial \mathbf{u}^2}$ is positive definite,
\item Co-State Dynamics:
\begin{displaymath}
\dot{\boldsymbol{\lambda}}(t) = -\frac{\partial \mathcal{H}}{\partial \mathbf{x}} = -\boldsymbol{\lambda}^T(t)\frac{\partial \mathbf{g}(\mathbf{x}, \mathbf{u})}{\partial \mathbf{x}} - \frac{\partial f(\mathbf{x}, \mathbf{u})}{\partial \mathbf{x}},
\end{displaymath}
\item State Dynamics: $\dot{\mathbf{x}}(t) = \frac{\partial \mathcal{H}}{\partial \boldsymbol{\lambda}} = \mathbf{g}(\mathbf{x}, \mathbf{u})$,
\item Initial Condition: $\mathbf{x}(0) = \mathbf{x}_0$, and 
\item Transversality Condition: $\boldsymbol{\lambda}(t_f) = \frac{\partial \Psi}{\partial \mathbf{x}}(\mathbf{x}(t_f))$.
\end{enumerate}
\label{lem:OptConNec}
\end{lemma}
We note that if a terminal condition $\mathbf{x}(t_f)$ is given, the transversality condition can be ignored in the necessary conditions.

%

\bibliography{AgentsConcealed,InfoGeom-2}

\begin{thebibliography}{105}%
\makeatletter
\providecommand \@ifxundefined [1]{%
 \@ifx{#1\undefined}
}%
\providecommand \@ifnum [1]{%
 \ifnum #1\expandafter \@firstoftwo
 \else \expandafter \@secondoftwo
 \fi
}%
\providecommand \@ifx [1]{%
 \ifx #1\expandafter \@firstoftwo
 \else \expandafter \@secondoftwo
 \fi
}%
\providecommand \natexlab [1]{#1}%
\providecommand \enquote  [1]{``#1''}%
\providecommand \bibnamefont  [1]{#1}%
\providecommand \bibfnamefont [1]{#1}%
\providecommand \citenamefont [1]{#1}%
\providecommand \href@noop [0]{\@secondoftwo}%
\providecommand \href [0]{\begingroup \@sanitize@url \@href}%
\providecommand \@href[1]{\@@startlink{#1}\@@href}%
\providecommand \@@href[1]{\endgroup#1\@@endlink}%
\providecommand \@sanitize@url [0]{\catcode `\\12\catcode `\$12\catcode
  `\&12\catcode `\#12\catcode `\^12\catcode `\_12\catcode `\%12\relax}%
\providecommand \@@startlink[1]{}%
\providecommand \@@endlink[0]{}%
\providecommand \url  [0]{\begingroup\@sanitize@url \@url }%
\providecommand \@url [1]{\endgroup\@href {#1}{\urlprefix }}%
\providecommand \urlprefix  [0]{URL }%
\providecommand \Eprint [0]{\href }%
\providecommand \doibase [0]{https://doi.org/}%
\providecommand \selectlanguage [0]{\@gobble}%
\providecommand \bibinfo  [0]{\@secondoftwo}%
\providecommand \bibfield  [0]{\@secondoftwo}%
\providecommand \translation [1]{[#1]}%
\providecommand \BibitemOpen [0]{}%
\providecommand \bibitemStop [0]{}%
\providecommand \bibitemNoStop [0]{.\EOS\space}%
\providecommand \EOS [0]{\spacefactor3000\relax}%
\providecommand \BibitemShut  [1]{\csname bibitem#1\endcsname}%
\let\auto@bib@innerbib\@empty
\bibitem [{\citenamefont {O'keefe}(2015)}]{O15}%
  \BibitemOpen
  \bibfield  {author} {\bibinfo {author} {\bibfnamefont {D.~J.}\ \bibnamefont
  {O'keefe}},\ }\href@noop {} {\emph {\bibinfo {title} {Persuasion: Theory and
  research}}}\ (\bibinfo  {publisher} {Sage Publications},\ \bibinfo {year}
  {2015})\BibitemShut {NoStop}%
\bibitem [{\citenamefont {Cervin}\ and\ \citenamefont
  {Henderson}(1961)}]{CH61}%
  \BibitemOpen
  \bibfield  {author} {\bibinfo {author} {\bibfnamefont {V.~B.}\ \bibnamefont
  {Cervin}}\ and\ \bibinfo {author} {\bibfnamefont {G.}~\bibnamefont
  {Henderson}},\ }\bibfield  {title} {\bibinfo {title} {Statistical theory of
  persuasion.},\ }\href@noop {} {\bibfield  {journal} {\bibinfo  {journal}
  {Psychological Review}\ }\textbf {\bibinfo {volume} {68}},\ \bibinfo {pages}
  {157} (\bibinfo {year} {1961})}\BibitemShut {NoStop}%
\bibitem [{\citenamefont {Burgoon}\ \emph {et~al.}(1981)\citenamefont
  {Burgoon}, \citenamefont {Burgoon}, \citenamefont {Miller},\ and\
  \citenamefont {Sunnafrank}}]{BBMS81}%
  \BibitemOpen
  \bibfield  {author} {\bibinfo {author} {\bibfnamefont {J.~K.}\ \bibnamefont
  {Burgoon}}, \bibinfo {author} {\bibfnamefont {M.}~\bibnamefont {Burgoon}},
  \bibinfo {author} {\bibfnamefont {G.~R.}\ \bibnamefont {Miller}},\ and\
  \bibinfo {author} {\bibfnamefont {M.}~\bibnamefont {Sunnafrank}},\ }\bibfield
   {title} {\bibinfo {title} {Learning theory approaches to persuasion},\
  }\href@noop {} {\bibfield  {journal} {\bibinfo  {journal} {Human
  Communication Research}\ }\textbf {\bibinfo {volume} {7}},\ \bibinfo {pages}
  {161} (\bibinfo {year} {1981})}\BibitemShut {NoStop}%
\bibitem [{\citenamefont {Curtis}\ and\ \citenamefont {Smith}(2008)}]{CS08}%
  \BibitemOpen
  \bibfield  {author} {\bibinfo {author} {\bibfnamefont {J.~P.}\ \bibnamefont
  {Curtis}}\ and\ \bibinfo {author} {\bibfnamefont {F.~T.}\ \bibnamefont
  {Smith}},\ }\bibfield  {title} {\bibinfo {title} {The dynamics of
  persuasion},\ }\href@noop {} {\bibfield  {journal} {\bibinfo  {journal} {Int.
  J. Math. Models Methods Appl. Sci}\ }\textbf {\bibinfo {volume} {2}},\
  \bibinfo {pages} {115} (\bibinfo {year} {2008})}\BibitemShut {NoStop}%
\bibitem [{\citenamefont {DeGroot}(1974)}]{DeGroot74}%
  \BibitemOpen
  \bibfield  {author} {\bibinfo {author} {\bibfnamefont {M.~H.}\ \bibnamefont
  {DeGroot}},\ }\bibfield  {title} {\bibinfo {title} {Reaching a consensus},\
  }\href@noop {} {\bibfield  {journal} {\bibinfo  {journal} {J. American Stat.
  Association}\ }\textbf {\bibinfo {volume} {69}},\ \bibinfo {pages} {118}
  (\bibinfo {year} {1974})}\BibitemShut {NoStop}%
\bibitem [{\citenamefont {Krause}(2000)}]{Krause00}%
  \BibitemOpen
  \bibfield  {author} {\bibinfo {author} {\bibfnamefont {U.}~\bibnamefont
  {Krause}},\ }\bibfield  {title} {\bibinfo {title} {A discrete nonlinear and
  non-autonomous model of consensus formation},\ }in\ \href@noop {} {\emph
  {\bibinfo {booktitle} {In Communications in Difference Equations}}},\
  \bibinfo {editor} {edited by\ \bibinfo {editor} {\bibnamefont {Gordon}}\ and\
  \bibinfo {editor} {\bibnamefont {Breach}}}\ (\bibinfo {year} {2000})\ pp.\
  \bibinfo {pages} {227-- 236}\BibitemShut {NoStop}%
\bibitem [{\citenamefont {Hegselmann}\ and\ \citenamefont
  {Krause}(2002)}]{HK02}%
  \BibitemOpen
  \bibfield  {author} {\bibinfo {author} {\bibfnamefont {R.}~\bibnamefont
  {Hegselmann}}\ and\ \bibinfo {author} {\bibfnamefont {U.}~\bibnamefont
  {Krause}},\ }\bibfield  {title} {\bibinfo {title} {Opinion dynamics and
  bounded confidence: Models, analysis and simulation},\ }\href@noop {}
  {\bibfield  {journal} {\bibinfo  {journal} {J. Artificial Soc. Social
  Simul.}\ }\textbf {\bibinfo {volume} {5}} (\bibinfo {year}
  {2002})}\BibitemShut {NoStop}%
\bibitem [{\citenamefont {Ben-Naim}(2005)}]{BN05}%
  \BibitemOpen
  \bibfield  {author} {\bibinfo {author} {\bibfnamefont {E.}~\bibnamefont
  {Ben-Naim}},\ }\bibfield  {title} {\bibinfo {title} {Opinion dynamics: Rise
  and fall of political parties},\ }\href@noop {} {\bibfield  {journal}
  {\bibinfo  {journal} {Europhys. Lett.}\ }\textbf {\bibinfo {volume} {69}},\
  \bibinfo {pages} {671} (\bibinfo {year} {2005})}\BibitemShut {NoStop}%
\bibitem [{\citenamefont {Weisbuch}\ \emph {et~al.}(2005)\citenamefont
  {Weisbuch}, \citenamefont {Deffuant},\ and\ \citenamefont {Amblard}}]{WDA05}%
  \BibitemOpen
  \bibfield  {author} {\bibinfo {author} {\bibfnamefont {G.}~\bibnamefont
  {Weisbuch}}, \bibinfo {author} {\bibfnamefont {G.}~\bibnamefont {Deffuant}},\
  and\ \bibinfo {author} {\bibfnamefont {F.}~\bibnamefont {Amblard}},\
  }\bibfield  {title} {\bibinfo {title} {Persuasion dynamics},\ }\href@noop {}
  {\bibfield  {journal} {\bibinfo  {journal} {Physica A}\ }\textbf {\bibinfo
  {volume} {353}} (\bibinfo {year} {2005})}\BibitemShut {NoStop}%
\bibitem [{\citenamefont {Toscani}(2006)}]{Toscani06}%
  \BibitemOpen
  \bibfield  {author} {\bibinfo {author} {\bibfnamefont {G.}~\bibnamefont
  {Toscani}},\ }\bibfield  {title} {\bibinfo {title} {Kinetic models of opinion
  formation},\ }\href@noop {} {\bibfield  {journal} {\bibinfo  {journal}
  {Commun. Math. Sci.}\ }\textbf {\bibinfo {volume} {4}},\ \bibinfo {pages}
  {481} (\bibinfo {year} {2006})}\BibitemShut {NoStop}%
\bibitem [{\citenamefont {Weisbuch}(2006)}]{Weisb06}%
  \BibitemOpen
  \bibfield  {author} {\bibinfo {author} {\bibfnamefont {G.}~\bibnamefont
  {Weisbuch}},\ }\bibfield  {title} {\bibinfo {title} {Social opinion
  dynamics},\ }in\ \href@noop {} {\emph {\bibinfo {booktitle} {Econophysics and
  Sociophysics: Trends and Perspectives}}},\ \bibinfo {editor} {edited by\
  \bibinfo {editor} {\bibfnamefont {B.~K.}\ \bibnamefont {Chakrabarti}},
  \bibinfo {editor} {\bibfnamefont {A.}~\bibnamefont {Chakrabarti}},\ and\
  \bibinfo {editor} {\bibfnamefont {A.}~\bibnamefont {Chatterjee}}}\ (\bibinfo
  {publisher} {Wiley},\ \bibinfo {year} {2006})\ pp.\ \bibinfo {pages}
  {67--94}\BibitemShut {NoStop}%
\bibitem [{\citenamefont {Lorenz}(2007)}]{Lorenz07}%
  \BibitemOpen
  \bibfield  {author} {\bibinfo {author} {\bibfnamefont {J.}~\bibnamefont
  {Lorenz}},\ }\bibfield  {title} {\bibinfo {title} {Continuous opinion
  dynamics of multidimensional allocation problems under bounded confidence. a
  survey},\ }\href@noop {} {\bibfield  {journal} {\bibinfo  {journal}
  {Internat. J. Modern Phys. C}\ }\textbf {\bibinfo {volume} {18}},\ \bibinfo
  {pages} {1819} (\bibinfo {year} {2007})}\BibitemShut {NoStop}%
\bibitem [{\citenamefont {Blondel}\ \emph {et~al.}(2009)\citenamefont
  {Blondel}, \citenamefont {Hendrickx},\ and\ \citenamefont
  {Tsitsiklis}}]{BHT09}%
  \BibitemOpen
  \bibfield  {author} {\bibinfo {author} {\bibfnamefont {V.~D.}\ \bibnamefont
  {Blondel}}, \bibinfo {author} {\bibfnamefont {J.~M.}\ \bibnamefont
  {Hendrickx}},\ and\ \bibinfo {author} {\bibfnamefont {J.~N.}\ \bibnamefont
  {Tsitsiklis}},\ }\bibfield  {title} {\bibinfo {title} {On krause's
  multi-agent consensus model with state-dependent connectivity},\ }\href
  {https://doi.org/10.1109/TAC.2009.2031211} {\bibfield  {journal} {\bibinfo
  {journal} {IEEE Transactions on Automatic Control}\ }\textbf {\bibinfo
  {volume} {54}},\ \bibinfo {pages} {2586} (\bibinfo {year}
  {2009})}\BibitemShut {NoStop}%
\bibitem [{\citenamefont {Castellano}\ \emph {et~al.}(2009)\citenamefont
  {Castellano}, \citenamefont {Fortunato},\ and\ \citenamefont
  {Loreto}}]{CFL09}%
  \BibitemOpen
  \bibfield  {author} {\bibinfo {author} {\bibfnamefont {C.}~\bibnamefont
  {Castellano}}, \bibinfo {author} {\bibfnamefont {S.}~\bibnamefont
  {Fortunato}},\ and\ \bibinfo {author} {\bibfnamefont {V.}~\bibnamefont
  {Loreto}},\ }\bibfield  {title} {\bibinfo {title} {Statistical physics of
  social dynamics},\ }\href@noop {} {\bibfield  {journal} {\bibinfo  {journal}
  {Reviews of modern physics}\ }\textbf {\bibinfo {volume} {81}},\ \bibinfo
  {pages} {591} (\bibinfo {year} {2009})}\BibitemShut {NoStop}%
\bibitem [{\citenamefont {Kurz}\ and\ \citenamefont {Rambau}(2011)}]{KR11}%
  \BibitemOpen
  \bibfield  {author} {\bibinfo {author} {\bibfnamefont {S.}~\bibnamefont
  {Kurz}}\ and\ \bibinfo {author} {\bibfnamefont {J.}~\bibnamefont {Rambau}},\
  }\bibfield  {title} {\bibinfo {title} {On the hegselmann-krause conjecture in
  opinion dynamics},\ }\href@noop {} {\bibfield  {journal} {\bibinfo  {journal}
  {J. Difference Equ. Appl.}\ }\textbf {\bibinfo {volume} {17}},\ \bibinfo
  {pages} {859} (\bibinfo {year} {2011})}\BibitemShut {NoStop}%
\bibitem [{\citenamefont {Duering}\ \emph {et~al.}(2012)\citenamefont
  {Duering}, \citenamefont {Markowich}, \citenamefont {Pietschmann},\ and\
  \citenamefont {Wolfram}}]{DMPW12}%
  \BibitemOpen
  \bibfield  {author} {\bibinfo {author} {\bibfnamefont {B.}~\bibnamefont
  {Duering}}, \bibinfo {author} {\bibfnamefont {P.}~\bibnamefont {Markowich}},
  \bibinfo {author} {\bibfnamefont {J.~F.}\ \bibnamefont {Pietschmann}},\ and\
  \bibinfo {author} {\bibfnamefont {M.~T.}\ \bibnamefont {Wolfram}},\
  }\bibfield  {title} {\bibinfo {title} {Boltzmann and {F}okker-{P}lanck
  equations modelling opinion formation in the presence of strong leaders},\
  }\href@noop {} {\bibfield  {journal} {\bibinfo  {journal} {Proc. R. Soc.
  Lond. Ser. A}\ }\textbf {\bibinfo {volume} {465}} (\bibinfo {year}
  {2012})}\BibitemShut {NoStop}%
\bibitem [{\citenamefont {Canuto}\ \emph {et~al.}(2012)\citenamefont {Canuto},
  \citenamefont {Fagnani},\ and\ \citenamefont {Tilli}}]{CFT12}%
  \BibitemOpen
  \bibfield  {author} {\bibinfo {author} {\bibfnamefont {C.}~\bibnamefont
  {Canuto}}, \bibinfo {author} {\bibfnamefont {F.}~\bibnamefont {Fagnani}},\
  and\ \bibinfo {author} {\bibfnamefont {P.}~\bibnamefont {Tilli}},\ }\bibfield
   {title} {\bibinfo {title} {An eulerian approach to the analysis of krause's
  consensus models},\ }\href@noop {} {\bibfield  {journal} {\bibinfo  {journal}
  {SIAM J. Contr. and Opt.}\ ,\ \bibinfo {pages} {243}} (\bibinfo {year}
  {2012})}\BibitemShut {NoStop}%
\bibitem [{\citenamefont {Jabin}\ and\ \citenamefont {Motsch}(2014)}]{JM14}%
  \BibitemOpen
  \bibfield  {author} {\bibinfo {author} {\bibfnamefont {P.-E.}\ \bibnamefont
  {Jabin}}\ and\ \bibinfo {author} {\bibfnamefont {S.}~\bibnamefont {Motsch}},\
  }\bibfield  {title} {\bibinfo {title} {Clustering and asymptotic behavior in
  opinion formation},\ }\href
  {https://doi.org/https://doi.org/10.1016/j.jde.2014.08.005} {\bibfield
  {journal} {\bibinfo  {journal} {Journal of Differential Equations}\ }\textbf
  {\bibinfo {volume} {257}},\ \bibinfo {pages} {4165 } (\bibinfo {year}
  {2014})}\BibitemShut {NoStop}%
\bibitem [{\citenamefont {Shang}\ \emph {et~al.}(2021)\citenamefont {Shang},
  \citenamefont {Zhao}, \citenamefont {Ai},\ and\ \citenamefont {Su}}]{SZAS21}%
  \BibitemOpen
  \bibfield  {author} {\bibinfo {author} {\bibfnamefont {L.}~\bibnamefont
  {Shang}}, \bibinfo {author} {\bibfnamefont {M.}~\bibnamefont {Zhao}},
  \bibinfo {author} {\bibfnamefont {J.}~\bibnamefont {Ai}},\ and\ \bibinfo
  {author} {\bibfnamefont {Z.}~\bibnamefont {Su}},\ }\bibfield  {title}
  {\bibinfo {title} {Opinion evolution in the sznajd model on interdependent
  chains},\ }\href
  {https://doi.org/https://doi.org/10.1016/j.physa.2020.125558} {\bibfield
  {journal} {\bibinfo  {journal} {Physica A: Statistical Mechanics and its
  Applications}\ }\textbf {\bibinfo {volume} {565}},\ \bibinfo {pages} {125558}
  (\bibinfo {year} {2021})}\BibitemShut {NoStop}%
\bibitem [{\citenamefont {Glass}\ and\ \citenamefont {Glass}(2021)}]{GG21}%
  \BibitemOpen
  \bibfield  {author} {\bibinfo {author} {\bibfnamefont {C.~A.}\ \bibnamefont
  {Glass}}\ and\ \bibinfo {author} {\bibfnamefont {D.~H.}\ \bibnamefont
  {Glass}},\ }\bibfield  {title} {\bibinfo {title} {Opinion dynamics of social
  learning with a conflicting source},\ }\href
  {https://doi.org/https://doi.org/10.1016/j.physa.2020.125480} {\bibfield
  {journal} {\bibinfo  {journal} {Physica A: Statistical Mechanics and its
  Applications}\ }\textbf {\bibinfo {volume} {563}},\ \bibinfo {pages} {125480}
  (\bibinfo {year} {2021})}\BibitemShut {NoStop}%
\bibitem [{\citenamefont {Centola}\ and\ \citenamefont
  {Baronchelli}(2015)}]{Centola15}%
  \BibitemOpen
  \bibfield  {author} {\bibinfo {author} {\bibfnamefont {D.}~\bibnamefont
  {Centola}}\ and\ \bibinfo {author} {\bibfnamefont {A.}~\bibnamefont
  {Baronchelli}},\ }\bibfield  {title} {\bibinfo {title} {Flocks, herds, and
  schools: A quantitative theory of flocking},\ }\href@noop {} {\bibfield
  {journal} {\bibinfo  {journal} {Proceedings of the National Academy of
  Sciences}\ }\textbf {\bibinfo {volume} {112}},\ \bibinfo {pages} {1989}
  (\bibinfo {year} {2015})}\BibitemShut {NoStop}%
\bibitem [{\citenamefont {Toner}\ and\ \citenamefont {Tu}(1998)}]{TT98}%
  \BibitemOpen
  \bibfield  {author} {\bibinfo {author} {\bibfnamefont {J.}~\bibnamefont
  {Toner}}\ and\ \bibinfo {author} {\bibfnamefont {Y.}~\bibnamefont {Tu}},\
  }\bibfield  {title} {\bibinfo {title} {Flocks, herds, and schools: A
  quantitative theory of flocking},\ }\href@noop {} {\bibfield  {journal}
  {\bibinfo  {journal} {Physical Review E}\ }\textbf {\bibinfo {volume} {58}},\
  \bibinfo {pages} {4828} (\bibinfo {year} {1998})}\BibitemShut {NoStop}%
\bibitem [{\citenamefont {Cucker}\ and\ \citenamefont {Smale}(2007)}]{CS07}%
  \BibitemOpen
  \bibfield  {author} {\bibinfo {author} {\bibfnamefont {F.}~\bibnamefont
  {Cucker}}\ and\ \bibinfo {author} {\bibfnamefont {S.}~\bibnamefont {Smale}},\
  }\bibfield  {title} {\bibinfo {title} {Emergent behavior in flocks},\
  }\href@noop {} {\bibfield  {journal} {\bibinfo  {journal} {IEEE Transactions
  on Automatic Control}\ }\textbf {\bibinfo {volume} {52}},\ \bibinfo {pages}
  {852} (\bibinfo {year} {2007})}\BibitemShut {NoStop}%
\bibitem [{\citenamefont {Edelstein-Keshet}(2001)}]{EK01}%
  \BibitemOpen
  \bibfield  {author} {\bibinfo {author} {\bibfnamefont {L.}~\bibnamefont
  {Edelstein-Keshet}},\ }\bibfield  {title} {\bibinfo {title} {Mathematical
  models of swarming and social aggregation},\ }in\ \href@noop {} {\emph
  {\bibinfo {booktitle} {Proc. 2001 International Symposium on Nonlinear Theory
  and Its Applications (NOLTA 2001)}}}\ (\bibinfo {address} {Miyagi, Japan},\
  \bibinfo {year} {2001})\BibitemShut {NoStop}%
\bibitem [{\citenamefont {Li}(2008)}]{L08}%
  \BibitemOpen
  \bibfield  {author} {\bibinfo {author} {\bibfnamefont {W.}~\bibnamefont
  {Li}},\ }\bibfield  {title} {\bibinfo {title} {Stability analysis of swarms
  with general topology},\ }\href@noop {} {\bibfield  {journal} {\bibinfo
  {journal} {IEEE Trans. Systems, Man and Cybernetics, Part B}\ }\textbf
  {\bibinfo {volume} {38}},\ \bibinfo {pages} {1084} (\bibinfo {year}
  {2008})}\BibitemShut {NoStop}%
\bibitem [{\citenamefont {Li}\ and\ \citenamefont {Xiao}(2010)}]{LX10}%
  \BibitemOpen
  \bibfield  {author} {\bibinfo {author} {\bibfnamefont {X.}~\bibnamefont
  {Li}}\ and\ \bibinfo {author} {\bibfnamefont {J.}~\bibnamefont {Xiao}},\
  }\bibfield  {title} {\bibinfo {title} {Swarming in homogeneous environments:
  A social interaction based framework},\ }\href@noop {} {\bibfield  {journal}
  {\bibinfo  {journal} {J. Theoret. Biol.}\ }\textbf {\bibinfo {volume}
  {264}},\ \bibinfo {pages} {747} (\bibinfo {year} {2010})}\BibitemShut
  {NoStop}%
\bibitem [{\citenamefont {Degond}\ and\ \citenamefont {Motsch}(2011)}]{DM11}%
  \BibitemOpen
  \bibfield  {author} {\bibinfo {author} {\bibfnamefont {P.}~\bibnamefont
  {Degond}}\ and\ \bibinfo {author} {\bibfnamefont {S.}~\bibnamefont
  {Motsch}},\ }\bibfield  {title} {\bibinfo {title} {A macroscopic model for a
  system of swarming agents using curvature control},\ }\href@noop {}
  {\bibfield  {journal} {\bibinfo  {journal} {J. Stat. Phys.}\ }\textbf
  {\bibinfo {volume} {143}} (\bibinfo {year} {2011})}\BibitemShut {NoStop}%
\bibitem [{\citenamefont {Motsch}\ and\ \citenamefont {Tadmor}(2014)}]{MT14}%
  \BibitemOpen
  \bibfield  {author} {\bibinfo {author} {\bibfnamefont {S.}~\bibnamefont
  {Motsch}}\ and\ \bibinfo {author} {\bibfnamefont {E.}~\bibnamefont
  {Tadmor}},\ }\bibfield  {title} {\bibinfo {title} {{Heterophilious Dynamics
  Enhances Consensus}},\ }\href@noop {} {\bibfield  {journal} {\bibinfo
  {journal} {SIAM Review}\ }\textbf {\bibinfo {volume} {56}},\ \bibinfo {pages}
  {577} (\bibinfo {year} {2014})}\BibitemShut {NoStop}%
\bibitem [{\citenamefont {Griffin}\ \emph {et~al.}(2022)\citenamefont
  {Griffin}, \citenamefont {Squicciarini},\ and\ \citenamefont {Jia}}]{GSJ22}%
  \BibitemOpen
  \bibfield  {author} {\bibinfo {author} {\bibfnamefont {C.}~\bibnamefont
  {Griffin}}, \bibinfo {author} {\bibfnamefont {A.}~\bibnamefont
  {Squicciarini}},\ and\ \bibinfo {author} {\bibfnamefont {F.}~\bibnamefont
  {Jia}},\ }\bibfield  {title} {\bibinfo {title} {Consensus in complex networks
  with noisy agents and peer pressure},\ }\href@noop {} {\bibfield  {journal}
  {\bibinfo  {journal} {Physica A: Statistical Mechanics and its Applications}\
  }\textbf {\bibinfo {volume} {608}},\ \bibinfo {pages} {128263} (\bibinfo
  {year} {2022})}\BibitemShut {NoStop}%
\bibitem [{\citenamefont {Huang}\ \emph {et~al.}(2016)\citenamefont {Huang},
  \citenamefont {Zhang}, \citenamefont {Xu},\ and\ \citenamefont
  {Fu}}]{HZXF16}%
  \BibitemOpen
  \bibfield  {author} {\bibinfo {author} {\bibfnamefont {W.-M.}\ \bibnamefont
  {Huang}}, \bibinfo {author} {\bibfnamefont {L.-J.}\ \bibnamefont {Zhang}},
  \bibinfo {author} {\bibfnamefont {X.-J.}\ \bibnamefont {Xu}},\ and\ \bibinfo
  {author} {\bibfnamefont {X.}~\bibnamefont {Fu}},\ }\bibfield  {title}
  {\bibinfo {title} {Contagion on complex networks with persuasion},\
  }\href@noop {} {\bibfield  {journal} {\bibinfo  {journal} {Scientific
  reports}\ }\textbf {\bibinfo {volume} {6}},\ \bibinfo {pages} {23766}
  (\bibinfo {year} {2016})}\BibitemShut {NoStop}%
\bibitem [{\citenamefont {Ratcliff}\ and\ \citenamefont {McKoon}(2008)}]{RM08}%
  \BibitemOpen
  \bibfield  {author} {\bibinfo {author} {\bibfnamefont {R.}~\bibnamefont
  {Ratcliff}}\ and\ \bibinfo {author} {\bibfnamefont {G.}~\bibnamefont
  {McKoon}},\ }\bibfield  {title} {\bibinfo {title} {The diffusion decision
  model: theory and data for two-choice decision tasks},\ }\href@noop {}
  {\bibfield  {journal} {\bibinfo  {journal} {Neural computation}\ }\textbf
  {\bibinfo {volume} {20}},\ \bibinfo {pages} {873} (\bibinfo {year}
  {2008})}\BibitemShut {NoStop}%
\bibitem [{\citenamefont {Xie}\ \emph {et~al.}(2021)\citenamefont {Xie},
  \citenamefont {Zhong}, \citenamefont {Li},\ and\ \citenamefont
  {Lui}}]{XZLL21}%
  \BibitemOpen
  \bibfield  {author} {\bibinfo {author} {\bibfnamefont {H.}~\bibnamefont
  {Xie}}, \bibinfo {author} {\bibfnamefont {M.}~\bibnamefont {Zhong}}, \bibinfo
  {author} {\bibfnamefont {Y.}~\bibnamefont {Li}},\ and\ \bibinfo {author}
  {\bibfnamefont {J.~C.}\ \bibnamefont {Lui}},\ }\bibfield  {title} {\bibinfo
  {title} {Understanding persuasion cascades in online product rating systems:
  Modeling, analysis, and inference},\ }\href@noop {} {\bibfield  {journal}
  {\bibinfo  {journal} {ACM Transactions on Knowledge Discovery from Data
  (TKDD)}\ }\textbf {\bibinfo {volume} {15}},\ \bibinfo {pages} {1} (\bibinfo
  {year} {2021})}\BibitemShut {NoStop}%
\bibitem [{\citenamefont {Kamenica}\ and\ \citenamefont
  {Gentzkow}(2011)}]{KG11}%
  \BibitemOpen
  \bibfield  {author} {\bibinfo {author} {\bibfnamefont {E.}~\bibnamefont
  {Kamenica}}\ and\ \bibinfo {author} {\bibfnamefont {M.}~\bibnamefont
  {Gentzkow}},\ }\bibfield  {title} {\bibinfo {title} {Bayesian persuasion},\
  }\href@noop {} {\bibfield  {journal} {\bibinfo  {journal} {American Economic
  Review}\ }\textbf {\bibinfo {volume} {101}},\ \bibinfo {pages} {2590}
  (\bibinfo {year} {2011})}\BibitemShut {NoStop}%
\bibitem [{\citenamefont {Babichenko}\ \emph {et~al.}(2021)\citenamefont
  {Babichenko}, \citenamefont {Talgam-Cohen},\ and\ \citenamefont
  {Zabarnyi}}]{BTZ21}%
  \BibitemOpen
  \bibfield  {author} {\bibinfo {author} {\bibfnamefont {Y.}~\bibnamefont
  {Babichenko}}, \bibinfo {author} {\bibfnamefont {I.}~\bibnamefont
  {Talgam-Cohen}},\ and\ \bibinfo {author} {\bibfnamefont {K.}~\bibnamefont
  {Zabarnyi}},\ }\bibfield  {title} {\bibinfo {title} {Bayesian persuasion
  under ex ante and ex post constraints},\ }in\ \href@noop {} {\emph {\bibinfo
  {booktitle} {Proceedings of the AAAI Conference on Artificial
  Intelligence}}},\ Vol.~\bibinfo {volume} {35}\ (\bibinfo {year} {2021})\ pp.\
  \bibinfo {pages} {5127--5134}\BibitemShut {NoStop}%
\bibitem [{\citenamefont {Caballero}\ and\ \citenamefont
  {Lunday}(2019)}]{CL19}%
  \BibitemOpen
  \bibfield  {author} {\bibinfo {author} {\bibfnamefont {W.~N.}\ \bibnamefont
  {Caballero}}\ and\ \bibinfo {author} {\bibfnamefont {B.~J.}\ \bibnamefont
  {Lunday}},\ }\bibfield  {title} {\bibinfo {title} {Influence modeling:
  Mathematical programming representations of persuasion under either risk or
  uncertainty},\ }\href@noop {} {\bibfield  {journal} {\bibinfo  {journal}
  {European Journal of Operational Research}\ }\textbf {\bibinfo {volume}
  {278}},\ \bibinfo {pages} {266} (\bibinfo {year} {2019})}\BibitemShut
  {NoStop}%
\bibitem [{\citenamefont {Altay}\ \emph {et~al.}(2023)\citenamefont {Altay},
  \citenamefont {Berriche},\ and\ \citenamefont {Acerbi}}]{ABA23}%
  \BibitemOpen
  \bibfield  {author} {\bibinfo {author} {\bibfnamefont {S.}~\bibnamefont
  {Altay}}, \bibinfo {author} {\bibfnamefont {M.}~\bibnamefont {Berriche}},\
  and\ \bibinfo {author} {\bibfnamefont {A.}~\bibnamefont {Acerbi}},\
  }\bibfield  {title} {\bibinfo {title} {Misinformation on misinformation:
  Conceptual and methodological challenges},\ }\href@noop {} {\bibfield
  {journal} {\bibinfo  {journal} {Social Media+ Society}\ }\textbf {\bibinfo
  {volume} {9}},\ \bibinfo {pages} {20563051221150412} (\bibinfo {year}
  {2023})}\BibitemShut {NoStop}%
\bibitem [{\citenamefont {Del~Vicario}\ \emph {et~al.}(2016)\citenamefont
  {Del~Vicario}, \citenamefont {Bessi}, \citenamefont {Zollo}, \citenamefont
  {Petroni}, \citenamefont {Scala}, \citenamefont {Caldarelli}, \citenamefont
  {Stanley},\ and\ \citenamefont {Quattrociocchi}}]{DBZP16}%
  \BibitemOpen
  \bibfield  {author} {\bibinfo {author} {\bibfnamefont {M.}~\bibnamefont
  {Del~Vicario}}, \bibinfo {author} {\bibfnamefont {A.}~\bibnamefont {Bessi}},
  \bibinfo {author} {\bibfnamefont {F.}~\bibnamefont {Zollo}}, \bibinfo
  {author} {\bibfnamefont {F.}~\bibnamefont {Petroni}}, \bibinfo {author}
  {\bibfnamefont {A.}~\bibnamefont {Scala}}, \bibinfo {author} {\bibfnamefont
  {G.}~\bibnamefont {Caldarelli}}, \bibinfo {author} {\bibfnamefont {H.~E.}\
  \bibnamefont {Stanley}},\ and\ \bibinfo {author} {\bibfnamefont
  {W.}~\bibnamefont {Quattrociocchi}},\ }\bibfield  {title} {\bibinfo {title}
  {The spreading of misinformation online},\ }\href@noop {} {\bibfield
  {journal} {\bibinfo  {journal} {Proceedings of the national academy of
  Sciences}\ }\textbf {\bibinfo {volume} {113}},\ \bibinfo {pages} {554}
  (\bibinfo {year} {2016})}\BibitemShut {NoStop}%
\bibitem [{\citenamefont {Vraga}\ and\ \citenamefont {Bode}(2020)}]{VB20}%
  \BibitemOpen
  \bibfield  {author} {\bibinfo {author} {\bibfnamefont {E.~K.}\ \bibnamefont
  {Vraga}}\ and\ \bibinfo {author} {\bibfnamefont {L.}~\bibnamefont {Bode}},\
  }\bibfield  {title} {\bibinfo {title} {Defining misinformation and
  understanding its bounded nature: Using expertise and evidence for describing
  misinformation},\ }\href@noop {} {\bibfield  {journal} {\bibinfo  {journal}
  {Political Communication}\ }\textbf {\bibinfo {volume} {37}},\ \bibinfo
  {pages} {136} (\bibinfo {year} {2020})}\BibitemShut {NoStop}%
\bibitem [{\citenamefont {Edelman}(2001)}]{E01}%
  \BibitemOpen
  \bibfield  {author} {\bibinfo {author} {\bibfnamefont {M.}~\bibnamefont
  {Edelman}},\ }\href@noop {} {\emph {\bibinfo {title} {The politics of
  misinformation}}}\ (\bibinfo  {publisher} {Cambridge University Press},\
  \bibinfo {year} {2001})\BibitemShut {NoStop}%
\bibitem [{\citenamefont {Jerit}\ and\ \citenamefont {Zhao}(2020)}]{JZ20}%
  \BibitemOpen
  \bibfield  {author} {\bibinfo {author} {\bibfnamefont {J.}~\bibnamefont
  {Jerit}}\ and\ \bibinfo {author} {\bibfnamefont {Y.}~\bibnamefont {Zhao}},\
  }\bibfield  {title} {\bibinfo {title} {Political misinformation},\
  }\href@noop {} {\bibfield  {journal} {\bibinfo  {journal} {Annual Review of
  Political Science}\ }\textbf {\bibinfo {volume} {23}},\ \bibinfo {pages} {77}
  (\bibinfo {year} {2020})}\BibitemShut {NoStop}%
\bibitem [{\citenamefont {Swire-Thompson}\ \emph {et~al.}(2020)\citenamefont
  {Swire-Thompson}, \citenamefont {Lazer} \emph {et~al.}}]{SLo20}%
  \BibitemOpen
  \bibfield  {author} {\bibinfo {author} {\bibfnamefont {B.}~\bibnamefont
  {Swire-Thompson}}, \bibinfo {author} {\bibfnamefont {D.}~\bibnamefont
  {Lazer}}, \emph {et~al.},\ }\bibfield  {title} {\bibinfo {title} {Public
  health and online misinformation: challenges and recommendations},\
  }\href@noop {} {\bibfield  {journal} {\bibinfo  {journal} {Annu Rev Public
  Health}\ }\textbf {\bibinfo {volume} {41}},\ \bibinfo {pages} {433} (\bibinfo
  {year} {2020})}\BibitemShut {NoStop}%
\bibitem [{\citenamefont {Southwell}\ \emph {et~al.}(2019)\citenamefont
  {Southwell}, \citenamefont {Niederdeppe}, \citenamefont {Cappella},
  \citenamefont {Gaysynsky}, \citenamefont {Kelley}, \citenamefont {Oh},
  \citenamefont {Peterson},\ and\ \citenamefont {Chou}}]{SNCG19}%
  \BibitemOpen
  \bibfield  {author} {\bibinfo {author} {\bibfnamefont {B.~G.}\ \bibnamefont
  {Southwell}}, \bibinfo {author} {\bibfnamefont {J.}~\bibnamefont
  {Niederdeppe}}, \bibinfo {author} {\bibfnamefont {J.~N.}\ \bibnamefont
  {Cappella}}, \bibinfo {author} {\bibfnamefont {A.}~\bibnamefont {Gaysynsky}},
  \bibinfo {author} {\bibfnamefont {D.~E.}\ \bibnamefont {Kelley}}, \bibinfo
  {author} {\bibfnamefont {A.}~\bibnamefont {Oh}}, \bibinfo {author}
  {\bibfnamefont {E.~B.}\ \bibnamefont {Peterson}},\ and\ \bibinfo {author}
  {\bibfnamefont {W.-Y.~S.}\ \bibnamefont {Chou}},\ }\bibfield  {title}
  {\bibinfo {title} {Misinformation as a misunderstood challenge to public
  health},\ }\href@noop {} {\bibfield  {journal} {\bibinfo  {journal} {American
  journal of preventive medicine}\ }\textbf {\bibinfo {volume} {57}},\ \bibinfo
  {pages} {282} (\bibinfo {year} {2019})}\BibitemShut {NoStop}%
\bibitem [{\citenamefont {Roozenbeek}\ \emph {et~al.}(2020)\citenamefont
  {Roozenbeek}, \citenamefont {Schneider}, \citenamefont {Dryhurst},
  \citenamefont {Kerr}, \citenamefont {Freeman}, \citenamefont {Recchia},
  \citenamefont {Van Der~Bles},\ and\ \citenamefont {Van Der~Linden}}]{RSDK20}%
  \BibitemOpen
  \bibfield  {author} {\bibinfo {author} {\bibfnamefont {J.}~\bibnamefont
  {Roozenbeek}}, \bibinfo {author} {\bibfnamefont {C.~R.}\ \bibnamefont
  {Schneider}}, \bibinfo {author} {\bibfnamefont {S.}~\bibnamefont {Dryhurst}},
  \bibinfo {author} {\bibfnamefont {J.}~\bibnamefont {Kerr}}, \bibinfo {author}
  {\bibfnamefont {A.~L.}\ \bibnamefont {Freeman}}, \bibinfo {author}
  {\bibfnamefont {G.}~\bibnamefont {Recchia}}, \bibinfo {author} {\bibfnamefont
  {A.~M.}\ \bibnamefont {Van Der~Bles}},\ and\ \bibinfo {author} {\bibfnamefont
  {S.}~\bibnamefont {Van Der~Linden}},\ }\bibfield  {title} {\bibinfo {title}
  {Susceptibility to misinformation about covid-19 around the world},\
  }\href@noop {} {\bibfield  {journal} {\bibinfo  {journal} {Royal Society open
  science}\ }\textbf {\bibinfo {volume} {7}},\ \bibinfo {pages} {201199}
  (\bibinfo {year} {2020})}\BibitemShut {NoStop}%
\bibitem [{\citenamefont {Joseph}\ \emph {et~al.}(2022)\citenamefont {Joseph},
  \citenamefont {Fernandez}, \citenamefont {Kritzman}, \citenamefont {Eaddy},
  \citenamefont {Cook}, \citenamefont {Lambros}, \citenamefont {Silva},
  \citenamefont {Arguelles}, \citenamefont {Abraham}, \citenamefont {Dorgham}
  \emph {et~al.}}]{JFKE22}%
  \BibitemOpen
  \bibfield  {author} {\bibinfo {author} {\bibfnamefont {A.~M.}\ \bibnamefont
  {Joseph}}, \bibinfo {author} {\bibfnamefont {V.}~\bibnamefont {Fernandez}},
  \bibinfo {author} {\bibfnamefont {S.}~\bibnamefont {Kritzman}}, \bibinfo
  {author} {\bibfnamefont {I.}~\bibnamefont {Eaddy}}, \bibinfo {author}
  {\bibfnamefont {O.~M.}\ \bibnamefont {Cook}}, \bibinfo {author}
  {\bibfnamefont {S.}~\bibnamefont {Lambros}}, \bibinfo {author} {\bibfnamefont
  {C.~E.~J.}\ \bibnamefont {Silva}}, \bibinfo {author} {\bibfnamefont
  {D.}~\bibnamefont {Arguelles}}, \bibinfo {author} {\bibfnamefont
  {C.}~\bibnamefont {Abraham}}, \bibinfo {author} {\bibfnamefont
  {N.}~\bibnamefont {Dorgham}}, \emph {et~al.},\ }\bibfield  {title} {\bibinfo
  {title} {Covid-19 misinformation on social media: a scoping review},\
  }\href@noop {} {\bibfield  {journal} {\bibinfo  {journal} {Cureus}\ }\textbf
  {\bibinfo {volume} {14}} (\bibinfo {year} {2022})}\BibitemShut {NoStop}%
\bibitem [{\citenamefont {Gisondi}\ \emph {et~al.}(2022)\citenamefont
  {Gisondi}, \citenamefont {Barber}, \citenamefont {Faust}, \citenamefont
  {Raja}, \citenamefont {Strehlow}, \citenamefont {Westafer},\ and\
  \citenamefont {Gottlieb}}]{GBFR22}%
  \BibitemOpen
  \bibfield  {author} {\bibinfo {author} {\bibfnamefont {M.~A.}\ \bibnamefont
  {Gisondi}}, \bibinfo {author} {\bibfnamefont {R.}~\bibnamefont {Barber}},
  \bibinfo {author} {\bibfnamefont {J.~S.}\ \bibnamefont {Faust}}, \bibinfo
  {author} {\bibfnamefont {A.}~\bibnamefont {Raja}}, \bibinfo {author}
  {\bibfnamefont {M.~C.}\ \bibnamefont {Strehlow}}, \bibinfo {author}
  {\bibfnamefont {L.~M.}\ \bibnamefont {Westafer}},\ and\ \bibinfo {author}
  {\bibfnamefont {M.}~\bibnamefont {Gottlieb}},\ }\href@noop {} {\bibinfo
  {title} {A deadly infodemic: social media and the power of covid-19
  misinformation}} (\bibinfo {year} {2022})\BibitemShut {NoStop}%
\bibitem [{\citenamefont {Whitehead}\ \emph {et~al.}(2023)\citenamefont
  {Whitehead}, \citenamefont {French}, \citenamefont {Caldwell}, \citenamefont
  {Letley},\ and\ \citenamefont {Mounier-Jack}}]{WFCL23}%
  \BibitemOpen
  \bibfield  {author} {\bibinfo {author} {\bibfnamefont {H.~S.}\ \bibnamefont
  {Whitehead}}, \bibinfo {author} {\bibfnamefont {C.~E.}\ \bibnamefont
  {French}}, \bibinfo {author} {\bibfnamefont {D.~M.}\ \bibnamefont
  {Caldwell}}, \bibinfo {author} {\bibfnamefont {L.}~\bibnamefont {Letley}},\
  and\ \bibinfo {author} {\bibfnamefont {S.}~\bibnamefont {Mounier-Jack}},\
  }\bibfield  {title} {\bibinfo {title} {A systematic review of communication
  interventions for countering vaccine misinformation},\ }\href@noop {}
  {\bibfield  {journal} {\bibinfo  {journal} {Vaccine}\ } (\bibinfo {year}
  {2023})}\BibitemShut {NoStop}%
\bibitem [{\citenamefont {Neely}\ \emph {et~al.}(2022)\citenamefont {Neely},
  \citenamefont {Eldredge}, \citenamefont {Ersing},\ and\ \citenamefont
  {Remington}}]{NEER22}%
  \BibitemOpen
  \bibfield  {author} {\bibinfo {author} {\bibfnamefont {S.~R.}\ \bibnamefont
  {Neely}}, \bibinfo {author} {\bibfnamefont {C.}~\bibnamefont {Eldredge}},
  \bibinfo {author} {\bibfnamefont {R.}~\bibnamefont {Ersing}},\ and\ \bibinfo
  {author} {\bibfnamefont {C.}~\bibnamefont {Remington}},\ }\bibfield  {title}
  {\bibinfo {title} {Vaccine hesitancy and exposure to misinformation: a survey
  analysis},\ }\href@noop {} {\bibfield  {journal} {\bibinfo  {journal}
  {Journal of general internal medicine}\ ,\ \bibinfo {pages} {1}} (\bibinfo
  {year} {2022})}\BibitemShut {NoStop}%
\bibitem [{\citenamefont {Enders}\ \emph {et~al.}(2022)\citenamefont {Enders},
  \citenamefont {Uscinski}, \citenamefont {Klofstad},\ and\ \citenamefont
  {Stoler}}]{EUKS22}%
  \BibitemOpen
  \bibfield  {author} {\bibinfo {author} {\bibfnamefont {A.~M.}\ \bibnamefont
  {Enders}}, \bibinfo {author} {\bibfnamefont {J.}~\bibnamefont {Uscinski}},
  \bibinfo {author} {\bibfnamefont {C.}~\bibnamefont {Klofstad}},\ and\
  \bibinfo {author} {\bibfnamefont {J.}~\bibnamefont {Stoler}},\ }\bibfield
  {title} {\bibinfo {title} {On the relationship between conspiracy theory
  beliefs, misinformation, and vaccine hesitancy},\ }\href@noop {} {\bibfield
  {journal} {\bibinfo  {journal} {Plos one}\ }\textbf {\bibinfo {volume}
  {17}},\ \bibinfo {pages} {e0276082} (\bibinfo {year} {2022})}\BibitemShut
  {NoStop}%
\bibitem [{\citenamefont {Cook}(2022)}]{C22}%
  \BibitemOpen
  \bibfield  {author} {\bibinfo {author} {\bibfnamefont {J.}~\bibnamefont
  {Cook}},\ }\bibfield  {title} {\bibinfo {title} {Understanding and countering
  misinformation about climate change},\ }\href@noop {} {\bibfield  {journal}
  {\bibinfo  {journal} {Research Anthology on Environmental and Societal
  Impacts of Climate Change}\ ,\ \bibinfo {pages} {1633}} (\bibinfo {year}
  {2022})}\BibitemShut {NoStop}%
\bibitem [{\citenamefont {Zhou}\ and\ \citenamefont {Shen}(2022)}]{ZS22}%
  \BibitemOpen
  \bibfield  {author} {\bibinfo {author} {\bibfnamefont {Y.}~\bibnamefont
  {Zhou}}\ and\ \bibinfo {author} {\bibfnamefont {L.}~\bibnamefont {Shen}},\
  }\bibfield  {title} {\bibinfo {title} {Confirmation bias and the persistence
  of misinformation on climate change},\ }\href@noop {} {\bibfield  {journal}
  {\bibinfo  {journal} {Communication Research}\ }\textbf {\bibinfo {volume}
  {49}},\ \bibinfo {pages} {500} (\bibinfo {year} {2022})}\BibitemShut
  {NoStop}%
\bibitem [{\citenamefont {Freiling}\ and\ \citenamefont
  {Matthes}(2023)}]{FM23}%
  \BibitemOpen
  \bibfield  {author} {\bibinfo {author} {\bibfnamefont {I.}~\bibnamefont
  {Freiling}}\ and\ \bibinfo {author} {\bibfnamefont {J.}~\bibnamefont
  {Matthes}},\ }\bibfield  {title} {\bibinfo {title} {Correcting climate change
  misinformation on social media: Reciprocal relationships between correcting
  others, anger, and environmental activism},\ }\href@noop {} {\bibfield
  {journal} {\bibinfo  {journal} {Computers in Human Behavior}\ }\textbf
  {\bibinfo {volume} {145}},\ \bibinfo {pages} {107769} (\bibinfo {year}
  {2023})}\BibitemShut {NoStop}%
\bibitem [{\citenamefont {Cook}\ \emph {et~al.}(2015)\citenamefont {Cook},
  \citenamefont {Ecker},\ and\ \citenamefont {Lewandowsky}}]{CEL15}%
  \BibitemOpen
  \bibfield  {author} {\bibinfo {author} {\bibfnamefont {J.}~\bibnamefont
  {Cook}}, \bibinfo {author} {\bibfnamefont {U.}~\bibnamefont {Ecker}},\ and\
  \bibinfo {author} {\bibfnamefont {S.}~\bibnamefont {Lewandowsky}},\
  }\bibfield  {title} {\bibinfo {title} {Misinformation and how to correct
  it},\ }\href@noop {} {\bibfield  {journal} {\bibinfo  {journal} {Emerging
  trends in the social and behavioral sciences: An interdisciplinary,
  searchable, and linkable resource}\ ,\ \bibinfo {pages} {1}} (\bibinfo {year}
  {2015})}\BibitemShut {NoStop}%
\bibitem [{\citenamefont {Van~der Linden}\ \emph
  {et~al.}(2017{\natexlab{a}})\citenamefont {Van~der Linden}, \citenamefont
  {Maibach}, \citenamefont {Cook}, \citenamefont {Leiserowitz},\ and\
  \citenamefont {Lewandowsky}}]{VMCL17}%
  \BibitemOpen
  \bibfield  {author} {\bibinfo {author} {\bibfnamefont {S.}~\bibnamefont
  {Van~der Linden}}, \bibinfo {author} {\bibfnamefont {E.}~\bibnamefont
  {Maibach}}, \bibinfo {author} {\bibfnamefont {J.}~\bibnamefont {Cook}},
  \bibinfo {author} {\bibfnamefont {A.}~\bibnamefont {Leiserowitz}},\ and\
  \bibinfo {author} {\bibfnamefont {S.}~\bibnamefont {Lewandowsky}},\
  }\bibfield  {title} {\bibinfo {title} {Inoculating against misinformation},\
  }\href@noop {} {\bibfield  {journal} {\bibinfo  {journal} {Science}\ }\textbf
  {\bibinfo {volume} {358}},\ \bibinfo {pages} {1141} (\bibinfo {year}
  {2017}{\natexlab{a}})}\BibitemShut {NoStop}%
\bibitem [{\citenamefont {Van~der Linden}\ \emph
  {et~al.}(2017{\natexlab{b}})\citenamefont {Van~der Linden}, \citenamefont
  {Leiserowitz}, \citenamefont {Rosenthal},\ and\ \citenamefont
  {Maibach}}]{VLRM17}%
  \BibitemOpen
  \bibfield  {author} {\bibinfo {author} {\bibfnamefont {S.}~\bibnamefont
  {Van~der Linden}}, \bibinfo {author} {\bibfnamefont {A.}~\bibnamefont
  {Leiserowitz}}, \bibinfo {author} {\bibfnamefont {S.}~\bibnamefont
  {Rosenthal}},\ and\ \bibinfo {author} {\bibfnamefont {E.}~\bibnamefont
  {Maibach}},\ }\bibfield  {title} {\bibinfo {title} {Inoculating the public
  against misinformation about climate change},\ }\href@noop {} {\bibfield
  {journal} {\bibinfo  {journal} {Global challenges}\ }\textbf {\bibinfo
  {volume} {1}},\ \bibinfo {pages} {1600008} (\bibinfo {year}
  {2017}{\natexlab{b}})}\BibitemShut {NoStop}%
\bibitem [{\citenamefont {Tay}\ \emph {et~al.}(2022)\citenamefont {Tay},
  \citenamefont {Hurlstone}, \citenamefont {Kurz},\ and\ \citenamefont
  {Ecker}}]{THKE22}%
  \BibitemOpen
  \bibfield  {author} {\bibinfo {author} {\bibfnamefont {L.~Q.}\ \bibnamefont
  {Tay}}, \bibinfo {author} {\bibfnamefont {M.~J.}\ \bibnamefont {Hurlstone}},
  \bibinfo {author} {\bibfnamefont {T.}~\bibnamefont {Kurz}},\ and\ \bibinfo
  {author} {\bibfnamefont {U.~K.}\ \bibnamefont {Ecker}},\ }\bibfield  {title}
  {\bibinfo {title} {A comparison of prebunking and debunking interventions for
  implied versus explicit misinformation},\ }\href@noop {} {\bibfield
  {journal} {\bibinfo  {journal} {British Journal of Psychology}\ }\textbf
  {\bibinfo {volume} {113}},\ \bibinfo {pages} {591} (\bibinfo {year}
  {2022})}\BibitemShut {NoStop}%
\bibitem [{\citenamefont {Ecker}\ \emph {et~al.}(2023)\citenamefont {Ecker},
  \citenamefont {Sharkey},\ and\ \citenamefont {Swire-Thompson}}]{ESS23}%
  \BibitemOpen
  \bibfield  {author} {\bibinfo {author} {\bibfnamefont {U.~K.}\ \bibnamefont
  {Ecker}}, \bibinfo {author} {\bibfnamefont {C.~X.}\ \bibnamefont {Sharkey}},\
  and\ \bibinfo {author} {\bibfnamefont {B.}~\bibnamefont {Swire-Thompson}},\
  }\bibfield  {title} {\bibinfo {title} {Correcting vaccine misinformation: A
  failure to replicate familiarity or fear-driven backfire effects},\
  }\href@noop {} {\bibfield  {journal} {\bibinfo  {journal} {Plos one}\
  }\textbf {\bibinfo {volume} {18}},\ \bibinfo {pages} {e0281140} (\bibinfo
  {year} {2023})}\BibitemShut {NoStop}%
\bibitem [{\citenamefont {Schmid-Petri}\ and\ \citenamefont
  {B{\"u}rger}(2022)}]{SB22}%
  \BibitemOpen
  \bibfield  {author} {\bibinfo {author} {\bibfnamefont {H.}~\bibnamefont
  {Schmid-Petri}}\ and\ \bibinfo {author} {\bibfnamefont {M.}~\bibnamefont
  {B{\"u}rger}},\ }\bibfield  {title} {\bibinfo {title} {The effect of
  misinformation and inoculation: Replication of an experiment on the effect of
  false experts in the context of climate change communication},\ }\href@noop
  {} {\bibfield  {journal} {\bibinfo  {journal} {Public Understanding of
  Science}\ }\textbf {\bibinfo {volume} {31}},\ \bibinfo {pages} {152}
  (\bibinfo {year} {2022})}\BibitemShut {NoStop}%
\bibitem [{\citenamefont {Buczel}\ \emph {et~al.}(2022)\citenamefont {Buczel},
  \citenamefont {Szyszka}, \citenamefont {Siwiak}, \citenamefont {Szpitalak},\
  and\ \citenamefont {Polczyk}}]{BSSS22}%
  \BibitemOpen
  \bibfield  {author} {\bibinfo {author} {\bibfnamefont {M.}~\bibnamefont
  {Buczel}}, \bibinfo {author} {\bibfnamefont {P.~D.}\ \bibnamefont {Szyszka}},
  \bibinfo {author} {\bibfnamefont {A.}~\bibnamefont {Siwiak}}, \bibinfo
  {author} {\bibfnamefont {M.}~\bibnamefont {Szpitalak}},\ and\ \bibinfo
  {author} {\bibfnamefont {R.}~\bibnamefont {Polczyk}},\ }\bibfield  {title}
  {\bibinfo {title} {Vaccination against misinformation: The inoculation
  technique reduces the continued influence effect},\ }\href@noop {} {\bibfield
   {journal} {\bibinfo  {journal} {Plos one}\ }\textbf {\bibinfo {volume}
  {17}},\ \bibinfo {pages} {e0267463} (\bibinfo {year} {2022})}\BibitemShut
  {NoStop}%
\bibitem [{\citenamefont {Faisal}\ and\ \citenamefont {Mahendra}(2022)}]{FM22}%
  \BibitemOpen
  \bibfield  {author} {\bibinfo {author} {\bibfnamefont {D.~R.}\ \bibnamefont
  {Faisal}}\ and\ \bibinfo {author} {\bibfnamefont {R.}~\bibnamefont
  {Mahendra}},\ }\bibfield  {title} {\bibinfo {title} {Two-stage classifier for
  covid-19 misinformation detection using bert: a study on indonesian tweets},\
  }\href@noop {} {\bibfield  {journal} {\bibinfo  {journal} {arXiv preprint
  arXiv:2206.15359}\ } (\bibinfo {year} {2022})}\BibitemShut {NoStop}%
\bibitem [{\citenamefont {Shao}\ \emph {et~al.}(2022)\citenamefont {Shao},
  \citenamefont {Sun}, \citenamefont {Zhang}, \citenamefont {Jiang},
  \citenamefont {Ma},\ and\ \citenamefont {Li}}]{SSZJ22}%
  \BibitemOpen
  \bibfield  {author} {\bibinfo {author} {\bibfnamefont {Y.}~\bibnamefont
  {Shao}}, \bibinfo {author} {\bibfnamefont {J.}~\bibnamefont {Sun}}, \bibinfo
  {author} {\bibfnamefont {T.}~\bibnamefont {Zhang}}, \bibinfo {author}
  {\bibfnamefont {Y.}~\bibnamefont {Jiang}}, \bibinfo {author} {\bibfnamefont
  {J.}~\bibnamefont {Ma}},\ and\ \bibinfo {author} {\bibfnamefont
  {J.}~\bibnamefont {Li}},\ }\bibfield  {title} {\bibinfo {title} {Fake news
  detection based on multi-modal classifier ensemble},\ }in\ \href@noop {}
  {\emph {\bibinfo {booktitle} {Proceedings of the 1st International Workshop
  on Multimedia AI against Disinformation}}}\ (\bibinfo {year} {2022})\ pp.\
  \bibinfo {pages} {78--86}\BibitemShut {NoStop}%
\bibitem [{\citenamefont {Fenza}\ \emph {et~al.}(2023)\citenamefont {Fenza},
  \citenamefont {Gallo}, \citenamefont {Loia}, \citenamefont {Petrone},\ and\
  \citenamefont {Stanzione}}]{FGLP23}%
  \BibitemOpen
  \bibfield  {author} {\bibinfo {author} {\bibfnamefont {G.}~\bibnamefont
  {Fenza}}, \bibinfo {author} {\bibfnamefont {M.}~\bibnamefont {Gallo}},
  \bibinfo {author} {\bibfnamefont {V.}~\bibnamefont {Loia}}, \bibinfo {author}
  {\bibfnamefont {A.}~\bibnamefont {Petrone}},\ and\ \bibinfo {author}
  {\bibfnamefont {C.}~\bibnamefont {Stanzione}},\ }\bibfield  {title} {\bibinfo
  {title} {Concept-drift detection index based on fuzzy formal concept analysis
  for fake news classifiers},\ }\href@noop {} {\bibfield  {journal} {\bibinfo
  {journal} {Technological Forecasting and Social Change}\ }\textbf {\bibinfo
  {volume} {194}},\ \bibinfo {pages} {122640} (\bibinfo {year}
  {2023})}\BibitemShut {NoStop}%
\bibitem [{\citenamefont {Mosallanezhad}\ \emph {et~al.}(2022)\citenamefont
  {Mosallanezhad}, \citenamefont {Karami}, \citenamefont {Shu}, \citenamefont
  {Mancenido},\ and\ \citenamefont {Liu}}]{MKSM22}%
  \BibitemOpen
  \bibfield  {author} {\bibinfo {author} {\bibfnamefont {A.}~\bibnamefont
  {Mosallanezhad}}, \bibinfo {author} {\bibfnamefont {M.}~\bibnamefont
  {Karami}}, \bibinfo {author} {\bibfnamefont {K.}~\bibnamefont {Shu}},
  \bibinfo {author} {\bibfnamefont {M.~V.}\ \bibnamefont {Mancenido}},\ and\
  \bibinfo {author} {\bibfnamefont {H.}~\bibnamefont {Liu}},\ }\bibfield
  {title} {\bibinfo {title} {Domain adaptive fake news detection via
  reinforcement learning},\ }in\ \href@noop {} {\emph {\bibinfo {booktitle}
  {Proceedings of the ACM Web Conference 2022}}}\ (\bibinfo {year} {2022})\
  pp.\ \bibinfo {pages} {3632--3640}\BibitemShut {NoStop}%
\bibitem [{\citenamefont {Ecker}\ \emph {et~al.}(2022)\citenamefont {Ecker},
  \citenamefont {Lewandowsky}, \citenamefont {Cook}, \citenamefont {Schmid},
  \citenamefont {Fazio}, \citenamefont {Brashier}, \citenamefont {Kendeou},
  \citenamefont {Vraga},\ and\ \citenamefont {Amazeen}}]{ELCS22}%
  \BibitemOpen
  \bibfield  {author} {\bibinfo {author} {\bibfnamefont {U.~K.}\ \bibnamefont
  {Ecker}}, \bibinfo {author} {\bibfnamefont {S.}~\bibnamefont {Lewandowsky}},
  \bibinfo {author} {\bibfnamefont {J.}~\bibnamefont {Cook}}, \bibinfo {author}
  {\bibfnamefont {P.}~\bibnamefont {Schmid}}, \bibinfo {author} {\bibfnamefont
  {L.~K.}\ \bibnamefont {Fazio}}, \bibinfo {author} {\bibfnamefont
  {N.}~\bibnamefont {Brashier}}, \bibinfo {author} {\bibfnamefont
  {P.}~\bibnamefont {Kendeou}}, \bibinfo {author} {\bibfnamefont {E.~K.}\
  \bibnamefont {Vraga}},\ and\ \bibinfo {author} {\bibfnamefont {M.~A.}\
  \bibnamefont {Amazeen}},\ }\bibfield  {title} {\bibinfo {title} {The
  psychological drivers of misinformation belief and its resistance to
  correction},\ }\href@noop {} {\bibfield  {journal} {\bibinfo  {journal}
  {Nature Reviews Psychology}\ }\textbf {\bibinfo {volume} {1}},\ \bibinfo
  {pages} {13} (\bibinfo {year} {2022})}\BibitemShut {NoStop}%
\bibitem [{\citenamefont {Reynolds}(2020)}]{R20}%
  \BibitemOpen
  \bibfield  {author} {\bibinfo {author} {\bibfnamefont {R.~M.}\ \bibnamefont
  {Reynolds}},\ }\href@noop {} {\emph {\bibinfo {title} {Why Does
  Misinformation Persist? Cognitive Explanations of the Implicit Message
  Effect}}}\ (\bibinfo  {publisher} {Michigan State University},\ \bibinfo
  {year} {2020})\BibitemShut {NoStop}%
\bibitem [{\citenamefont {Walter}\ and\ \citenamefont
  {Tukachinsky}(2020)}]{WT20}%
  \BibitemOpen
  \bibfield  {author} {\bibinfo {author} {\bibfnamefont {N.}~\bibnamefont
  {Walter}}\ and\ \bibinfo {author} {\bibfnamefont {R.}~\bibnamefont
  {Tukachinsky}},\ }\bibfield  {title} {\bibinfo {title} {A meta-analytic
  examination of the continued influence of misinformation in the face of
  correction: How powerful is it, why does it happen, and how to stop it?},\
  }\href@noop {} {\bibfield  {journal} {\bibinfo  {journal} {Communication
  research}\ }\textbf {\bibinfo {volume} {47}},\ \bibinfo {pages} {155}
  (\bibinfo {year} {2020})}\BibitemShut {NoStop}%
\bibitem [{\citenamefont {Sindermann}\ \emph {et~al.}(2020)\citenamefont
  {Sindermann}, \citenamefont {Cooper},\ and\ \citenamefont {Montag}}]{SCM20}%
  \BibitemOpen
  \bibfield  {author} {\bibinfo {author} {\bibfnamefont {C.}~\bibnamefont
  {Sindermann}}, \bibinfo {author} {\bibfnamefont {A.}~\bibnamefont {Cooper}},\
  and\ \bibinfo {author} {\bibfnamefont {C.}~\bibnamefont {Montag}},\
  }\bibfield  {title} {\bibinfo {title} {A short review on susceptibility to
  falling for fake political news},\ }\href@noop {} {\bibfield  {journal}
  {\bibinfo  {journal} {Current Opinion in Psychology}\ }\textbf {\bibinfo
  {volume} {36}},\ \bibinfo {pages} {44} (\bibinfo {year} {2020})}\BibitemShut
  {NoStop}%
\bibitem [{\citenamefont {Chaxel}(2022)}]{C22a}%
  \BibitemOpen
  \bibfield  {author} {\bibinfo {author} {\bibfnamefont {A.-S.}\ \bibnamefont
  {Chaxel}},\ }\bibfield  {title} {\bibinfo {title} {How misinformation taints
  our belief system: A focus on belief updating and relational reasoning},\
  }\href@noop {} {\bibfield  {journal} {\bibinfo  {journal} {Journal of
  Consumer Psychology}\ }\textbf {\bibinfo {volume} {32}},\ \bibinfo {pages}
  {370} (\bibinfo {year} {2022})}\BibitemShut {NoStop}%
\bibitem [{\citenamefont {Blake}\ and\ \citenamefont {Mullin}(2014)}]{BM14}%
  \BibitemOpen
  \bibfield  {author} {\bibinfo {author} {\bibfnamefont {I.~F.}\ \bibnamefont
  {Blake}}\ and\ \bibinfo {author} {\bibfnamefont {R.~C.}\ \bibnamefont
  {Mullin}},\ }\href@noop {} {\emph {\bibinfo {title} {The mathematical theory
  of coding}}}\ (\bibinfo  {publisher} {Academic Press},\ \bibinfo {year}
  {2014})\BibitemShut {NoStop}%
\bibitem [{\citenamefont {Crooks}(2007)}]{C07}%
  \BibitemOpen
  \bibfield  {author} {\bibinfo {author} {\bibfnamefont {G.~E.}\ \bibnamefont
  {Crooks}},\ }\bibfield  {title} {\bibinfo {title} {Measuring thermodynamic
  length},\ }\href@noop {} {\bibfield  {journal} {\bibinfo  {journal} {Physical
  Review Letters}\ }\textbf {\bibinfo {volume} {99}},\ \bibinfo {pages}
  {100602} (\bibinfo {year} {2007})}\BibitemShut {NoStop}%
\bibitem [{\citenamefont {Bordel}(2011)}]{B11}%
  \BibitemOpen
  \bibfield  {author} {\bibinfo {author} {\bibfnamefont {S.}~\bibnamefont
  {Bordel}},\ }\bibfield  {title} {\bibinfo {title} {Non-equilibrium
  statistical mechanics: partition functions and steepest entropy increase},\
  }\href@noop {} {\bibfield  {journal} {\bibinfo  {journal} {Journal of
  Statistical Mechanics: Theory and Experiment}\ }\textbf {\bibinfo {volume}
  {2011}},\ \bibinfo {pages} {P05013} (\bibinfo {year} {2011})}\BibitemShut
  {NoStop}%
\bibitem [{\citenamefont {Still}\ \emph {et~al.}(2012)\citenamefont {Still},
  \citenamefont {Sivak}, \citenamefont {Bell},\ and\ \citenamefont
  {Crooks}}]{SSBC12}%
  \BibitemOpen
  \bibfield  {author} {\bibinfo {author} {\bibfnamefont {S.}~\bibnamefont
  {Still}}, \bibinfo {author} {\bibfnamefont {D.~A.}\ \bibnamefont {Sivak}},
  \bibinfo {author} {\bibfnamefont {A.~J.}\ \bibnamefont {Bell}},\ and\
  \bibinfo {author} {\bibfnamefont {G.~E.}\ \bibnamefont {Crooks}},\ }\bibfield
   {title} {\bibinfo {title} {Thermodynamics of prediction},\ }\href@noop {}
  {\bibfield  {journal} {\bibinfo  {journal} {Physical review letters}\
  }\textbf {\bibinfo {volume} {109}},\ \bibinfo {pages} {120604} (\bibinfo
  {year} {2012})}\BibitemShut {NoStop}%
\bibitem [{\citenamefont {Sivak}\ and\ \citenamefont {Crooks}(2012)}]{SC12}%
  \BibitemOpen
  \bibfield  {author} {\bibinfo {author} {\bibfnamefont {D.~A.}\ \bibnamefont
  {Sivak}}\ and\ \bibinfo {author} {\bibfnamefont {G.~E.}\ \bibnamefont
  {Crooks}},\ }\bibfield  {title} {\bibinfo {title} {Thermodynamic metrics and
  optimal paths},\ }\href@noop {} {\bibfield  {journal} {\bibinfo  {journal}
  {Physical review letters}\ }\textbf {\bibinfo {volume} {108}},\ \bibinfo
  {pages} {190602} (\bibinfo {year} {2012})}\BibitemShut {NoStop}%
\bibitem [{\citenamefont {Kim}\ \emph {et~al.}(2016)\citenamefont {Kim},
  \citenamefont {Lee}, \citenamefont {Heseltine},\ and\ \citenamefont
  {Hollerbach}}]{KLHH16}%
  \BibitemOpen
  \bibfield  {author} {\bibinfo {author} {\bibfnamefont {E.-j.}\ \bibnamefont
  {Kim}}, \bibinfo {author} {\bibfnamefont {U.}~\bibnamefont {Lee}}, \bibinfo
  {author} {\bibfnamefont {J.}~\bibnamefont {Heseltine}},\ and\ \bibinfo
  {author} {\bibfnamefont {R.}~\bibnamefont {Hollerbach}},\ }\bibfield  {title}
  {\bibinfo {title} {Geometric structure and geodesic in a solvable model of
  nonequilibrium process},\ }\href@noop {} {\bibfield  {journal} {\bibinfo
  {journal} {Physical Review E}\ }\textbf {\bibinfo {volume} {93}},\ \bibinfo
  {pages} {062127} (\bibinfo {year} {2016})}\BibitemShut {NoStop}%
\bibitem [{\citenamefont {Feng}\ and\ \citenamefont {Crooks}(2009)}]{FC09}%
  \BibitemOpen
  \bibfield  {author} {\bibinfo {author} {\bibfnamefont {E.~H.}\ \bibnamefont
  {Feng}}\ and\ \bibinfo {author} {\bibfnamefont {G.~E.}\ \bibnamefont
  {Crooks}},\ }\bibfield  {title} {\bibinfo {title} {Far-from-equilibrium
  measurements of thermodynamic length},\ }\href@noop {} {\bibfield  {journal}
  {\bibinfo  {journal} {Physical review E}\ }\textbf {\bibinfo {volume} {79}},\
  \bibinfo {pages} {012104} (\bibinfo {year} {2009})}\BibitemShut {NoStop}%
\bibitem [{\citenamefont {Kim}\ and\ \citenamefont {Hollerbach}(2017)}]{KH17}%
  \BibitemOpen
  \bibfield  {author} {\bibinfo {author} {\bibfnamefont {E.-j.}\ \bibnamefont
  {Kim}}\ and\ \bibinfo {author} {\bibfnamefont {R.}~\bibnamefont
  {Hollerbach}},\ }\bibfield  {title} {\bibinfo {title} {Geometric structure
  and information change in phase transitions},\ }\href@noop {} {\bibfield
  {journal} {\bibinfo  {journal} {Physical Review E}\ }\textbf {\bibinfo
  {volume} {95}},\ \bibinfo {pages} {062107} (\bibinfo {year}
  {2017})}\BibitemShut {NoStop}%
\bibitem [{\citenamefont {Kim}(2021)}]{K21}%
  \BibitemOpen
  \bibfield  {author} {\bibinfo {author} {\bibfnamefont {E.-j.}\ \bibnamefont
  {Kim}},\ }\bibfield  {title} {\bibinfo {title} {Information geometry and
  non-equilibrium thermodynamic relations in the over-damped stochastic
  processes},\ }\href@noop {} {\bibfield  {journal} {\bibinfo  {journal}
  {Journal of Statistical Mechanics: Theory and Experiment}\ }\textbf {\bibinfo
  {volume} {2021}},\ \bibinfo {pages} {093406} (\bibinfo {year}
  {2021})}\BibitemShut {NoStop}%
\bibitem [{\citenamefont {Gomez}\ \emph {et~al.}(2020)\citenamefont {Gomez},
  \citenamefont {Portesi},\ and\ \citenamefont {Borges}}]{GPB20}%
  \BibitemOpen
  \bibfield  {author} {\bibinfo {author} {\bibfnamefont {I.~S.}\ \bibnamefont
  {Gomez}}, \bibinfo {author} {\bibfnamefont {M.}~\bibnamefont {Portesi}},\
  and\ \bibinfo {author} {\bibfnamefont {E.~P.}\ \bibnamefont {Borges}},\
  }\bibfield  {title} {\bibinfo {title} {Universality classes for the fisher
  metric derived from relative group entropy},\ }\href@noop {} {\bibfield
  {journal} {\bibinfo  {journal} {Physica A: Statistical Mechanics and its
  Applications}\ }\textbf {\bibinfo {volume} {547}},\ \bibinfo {pages} {123827}
  (\bibinfo {year} {2020})}\BibitemShut {NoStop}%
\bibitem [{\citenamefont {Nicholson}\ \emph {et~al.}(2020)\citenamefont
  {Nicholson}, \citenamefont {Garc{\'\i}a-Pintos}, \citenamefont {del Campo},\
  and\ \citenamefont {Green}}]{NGCG20}%
  \BibitemOpen
  \bibfield  {author} {\bibinfo {author} {\bibfnamefont {S.~B.}\ \bibnamefont
  {Nicholson}}, \bibinfo {author} {\bibfnamefont {L.~P.}\ \bibnamefont
  {Garc{\'\i}a-Pintos}}, \bibinfo {author} {\bibfnamefont {A.}~\bibnamefont
  {del Campo}},\ and\ \bibinfo {author} {\bibfnamefont {J.~R.}\ \bibnamefont
  {Green}},\ }\bibfield  {title} {\bibinfo {title} {Time--information
  uncertainty relations in thermodynamics},\ }\href@noop {} {\bibfield
  {journal} {\bibinfo  {journal} {Nature Physics}\ }\textbf {\bibinfo {volume}
  {16}},\ \bibinfo {pages} {1211} (\bibinfo {year} {2020})}\BibitemShut
  {NoStop}%
\bibitem [{\citenamefont {Fujiwara}\ and\ \citenamefont {Amari}(1995)}]{FA95}%
  \BibitemOpen
  \bibfield  {author} {\bibinfo {author} {\bibfnamefont {A.}~\bibnamefont
  {Fujiwara}}\ and\ \bibinfo {author} {\bibfnamefont {S.-i.}\ \bibnamefont
  {Amari}},\ }\bibfield  {title} {\bibinfo {title} {Gradient systems in view of
  information geometry},\ }\href@noop {} {\bibfield  {journal} {\bibinfo
  {journal} {Physica D: Nonlinear Phenomena}\ }\textbf {\bibinfo {volume}
  {80}},\ \bibinfo {pages} {317} (\bibinfo {year} {1995})}\BibitemShut
  {NoStop}%
\bibitem [{\citenamefont {Zhang}\ \emph {et~al.}(2020)\citenamefont {Zhang},
  \citenamefont {Guan},\ and\ \citenamefont {Shi}}]{ZGS20}%
  \BibitemOpen
  \bibfield  {author} {\bibinfo {author} {\bibfnamefont {Z.}~\bibnamefont
  {Zhang}}, \bibinfo {author} {\bibfnamefont {S.}~\bibnamefont {Guan}},\ and\
  \bibinfo {author} {\bibfnamefont {H.}~\bibnamefont {Shi}},\ }\bibfield
  {title} {\bibinfo {title} {Information geometry in the population dynamics of
  bacteria},\ }\href@noop {} {\bibfield  {journal} {\bibinfo  {journal}
  {Journal of Statistical Mechanics: Theory and Experiment}\ }\textbf {\bibinfo
  {volume} {2020}},\ \bibinfo {pages} {073501} (\bibinfo {year}
  {2020})}\BibitemShut {NoStop}%
\bibitem [{\citenamefont {Polettini}\ and\ \citenamefont
  {Esposito}(2013)}]{PE13}%
  \BibitemOpen
  \bibfield  {author} {\bibinfo {author} {\bibfnamefont {M.}~\bibnamefont
  {Polettini}}\ and\ \bibinfo {author} {\bibfnamefont {M.}~\bibnamefont
  {Esposito}},\ }\bibfield  {title} {\bibinfo {title} {Nonconvexity of the
  relative entropy for markov dynamics: A fisher information approach},\
  }\href@noop {} {\bibfield  {journal} {\bibinfo  {journal} {Physical Review
  E}\ }\textbf {\bibinfo {volume} {88}},\ \bibinfo {pages} {012112} (\bibinfo
  {year} {2013})}\BibitemShut {NoStop}%
\bibitem [{\citenamefont {Amari}(1997)}]{A97}%
  \BibitemOpen
  \bibfield  {author} {\bibinfo {author} {\bibfnamefont {S.-i.}\ \bibnamefont
  {Amari}},\ }\bibfield  {title} {\bibinfo {title} {Information geometry of
  neural networks---an overview---},\ }\href@noop {} {\bibfield  {journal}
  {\bibinfo  {journal} {Mathematics of Neural Networks: Models, Algorithms and
  Applications}\ ,\ \bibinfo {pages} {15}} (\bibinfo {year}
  {1997})}\BibitemShut {NoStop}%
\bibitem [{\citenamefont {Ollivier}\ \emph {et~al.}(2017)\citenamefont
  {Ollivier}, \citenamefont {Arnold}, \citenamefont {Auger},\ and\
  \citenamefont {Hansen}}]{OAAH17}%
  \BibitemOpen
  \bibfield  {author} {\bibinfo {author} {\bibfnamefont {Y.}~\bibnamefont
  {Ollivier}}, \bibinfo {author} {\bibfnamefont {L.}~\bibnamefont {Arnold}},
  \bibinfo {author} {\bibfnamefont {A.}~\bibnamefont {Auger}},\ and\ \bibinfo
  {author} {\bibfnamefont {N.}~\bibnamefont {Hansen}},\ }\bibfield  {title}
  {\bibinfo {title} {Information-geometric optimization algorithms: A unifying
  picture via invariance principles},\ }\href@noop {} {\bibfield  {journal}
  {\bibinfo  {journal} {The Journal of Machine Learning Research}\ }\textbf
  {\bibinfo {volume} {18}},\ \bibinfo {pages} {564} (\bibinfo {year}
  {2017})}\BibitemShut {NoStop}%
\bibitem [{\citenamefont {Kullback}\ and\ \citenamefont
  {Leibler}(1951)}]{Kullback1951}%
  \BibitemOpen
  \bibfield  {author} {\bibinfo {author} {\bibfnamefont {S.}~\bibnamefont
  {Kullback}}\ and\ \bibinfo {author} {\bibfnamefont {R.~A.}\ \bibnamefont
  {Leibler}},\ }\bibfield  {title} {\bibinfo {title} {{On Information and
  Sufficiency}},\ }\href {https://doi.org/10.1214/aoms/1177729694} {\bibfield
  {journal} {\bibinfo  {journal} {The Annals of Mathematical Statistics}\
  }\textbf {\bibinfo {volume} {22}},\ \bibinfo {pages} {79 } (\bibinfo {year}
  {1951})}\BibitemShut {NoStop}%
\bibitem [{\citenamefont {Bishop}(2006)}]{bishop2006}%
  \BibitemOpen
  \bibfield  {author} {\bibinfo {author} {\bibfnamefont {C.~M.}\ \bibnamefont
  {Bishop}},\ }\href@noop {} {\emph {\bibinfo {title} {Pattern recognition and
  machine learning}}}\ (\bibinfo  {publisher} {Springer},\ \bibinfo {address}
  {New York, NY},\ \bibinfo {year} {2006})\BibitemShut {NoStop}%
\bibitem [{\citenamefont {Dowty}(2018)}]{D18}%
  \BibitemOpen
  \bibfield  {author} {\bibinfo {author} {\bibfnamefont {J.~G.}\ \bibnamefont
  {Dowty}},\ }\bibfield  {title} {\bibinfo {title} {Chentsov's theorem for
  exponential families},\ }\href@noop {} {\bibfield  {journal} {\bibinfo
  {journal} {Information Geometry}\ }\textbf {\bibinfo {volume} {1}},\ \bibinfo
  {pages} {117} (\bibinfo {year} {2018})}\BibitemShut {NoStop}%
\bibitem [{\citenamefont {Caticha}(2015)}]{C15}%
  \BibitemOpen
  \bibfield  {author} {\bibinfo {author} {\bibfnamefont {A.}~\bibnamefont
  {Caticha}},\ }\bibfield  {title} {\bibinfo {title} {The basics of information
  geometry},\ }in\ \href@noop {} {\emph {\bibinfo {booktitle} {AIP Conference
  Proceedings}}},\ Vol.\ \bibinfo {volume} {1641}\ (\bibinfo {organization}
  {American Institute of Physics},\ \bibinfo {year} {2015})\ pp.\ \bibinfo
  {pages} {15--26}\BibitemShut {NoStop}%
\bibitem [{\citenamefont {Kirk}(2004)}]{K04}%
  \BibitemOpen
  \bibfield  {author} {\bibinfo {author} {\bibfnamefont {D.~E.}\ \bibnamefont
  {Kirk}},\ }\href@noop {} {\emph {\bibinfo {title} {Optimal control theory: an
  introduction}}}\ (\bibinfo  {publisher} {Courier Corporation},\ \bibinfo
  {year} {2004})\BibitemShut {NoStop}%
\bibitem [{\citenamefont {Cover}\ and\ \citenamefont
  {Thomas}(2005)}]{cover2005}%
  \BibitemOpen
  \bibfield  {author} {\bibinfo {author} {\bibfnamefont {T.~M.}\ \bibnamefont
  {Cover}}\ and\ \bibinfo {author} {\bibfnamefont {J.~A.}\ \bibnamefont
  {Thomas}},\ }\href@noop {} {\emph {\bibinfo {title} {Elements of Information
  Theory}}}\ (\bibinfo  {publisher} {Wiley},\ \bibinfo {address} {Hoboken,
  NJ},\ \bibinfo {year} {2005})\BibitemShut {NoStop}%
\bibitem [{\citenamefont {Rao}\ and\ \citenamefont {Ballard}(1999)}]{RB99}%
  \BibitemOpen
  \bibfield  {author} {\bibinfo {author} {\bibfnamefont {R.~P.}\ \bibnamefont
  {Rao}}\ and\ \bibinfo {author} {\bibfnamefont {D.~H.}\ \bibnamefont
  {Ballard}},\ }\bibfield  {title} {\bibinfo {title} {Predictive coding in the
  visual cortex: a functional interpretation of some extra-classical
  receptive-field effects},\ }\href@noop {} {\bibfield  {journal} {\bibinfo
  {journal} {Nature neuroscience}\ }\textbf {\bibinfo {volume} {2}},\ \bibinfo
  {pages} {79} (\bibinfo {year} {1999})}\BibitemShut {NoStop}%
\bibitem [{\citenamefont {Friston}(2012)}]{F12}%
  \BibitemOpen
  \bibfield  {author} {\bibinfo {author} {\bibfnamefont {K.}~\bibnamefont
  {Friston}},\ }\bibfield  {title} {\bibinfo {title} {The history of the future
  of the bayesian brain},\ }\href@noop {} {\bibfield  {journal} {\bibinfo
  {journal} {NeuroImage}\ }\textbf {\bibinfo {volume} {62}},\ \bibinfo {pages}
  {1230} (\bibinfo {year} {2012})}\BibitemShut {NoStop}%
\bibitem [{\citenamefont {Baltieri}\ and\ \citenamefont
  {Isomura}(2021)}]{BI21}%
  \BibitemOpen
  \bibfield  {author} {\bibinfo {author} {\bibfnamefont {M.}~\bibnamefont
  {Baltieri}}\ and\ \bibinfo {author} {\bibfnamefont {T.}~\bibnamefont
  {Isomura}},\ }\bibfield  {title} {\bibinfo {title} {Kalman filters as the
  steady-state solution of gradient descent on variational free energy},\
  }\href@noop {} {\bibfield  {journal} {\bibinfo  {journal} {arXiv preprint
  arXiv:2111.10530}\ } (\bibinfo {year} {2021})}\BibitemShut {NoStop}%
\bibitem [{\citenamefont {Millidge}\ \emph {et~al.}(2021)\citenamefont
  {Millidge}, \citenamefont {Tschantz}, \citenamefont {Seth},\ and\
  \citenamefont {Buckley}}]{MTSB21}%
  \BibitemOpen
  \bibfield  {author} {\bibinfo {author} {\bibfnamefont {B.}~\bibnamefont
  {Millidge}}, \bibinfo {author} {\bibfnamefont {A.}~\bibnamefont {Tschantz}},
  \bibinfo {author} {\bibfnamefont {A.}~\bibnamefont {Seth}},\ and\ \bibinfo
  {author} {\bibfnamefont {C.}~\bibnamefont {Buckley}},\ }\bibfield  {title}
  {\bibinfo {title} {Neural kalman filtering},\ }\href@noop {} {\bibfield
  {journal} {\bibinfo  {journal} {arXiv preprint arXiv:2102.10021}\ } (\bibinfo
  {year} {2021})}\BibitemShut {NoStop}%
\bibitem [{\citenamefont {de~Xivry}\ \emph {et~al.}(2013)\citenamefont
  {de~Xivry}, \citenamefont {Coppe}, \citenamefont {Blohm},\ and\ \citenamefont
  {Lefevre}}]{XCBL13}%
  \BibitemOpen
  \bibfield  {author} {\bibinfo {author} {\bibfnamefont {J.-J.~O.}\
  \bibnamefont {de~Xivry}}, \bibinfo {author} {\bibfnamefont {S.}~\bibnamefont
  {Coppe}}, \bibinfo {author} {\bibfnamefont {G.}~\bibnamefont {Blohm}},\ and\
  \bibinfo {author} {\bibfnamefont {P.}~\bibnamefont {Lefevre}},\ }\bibfield
  {title} {\bibinfo {title} {Kalman filtering naturally accounts for visually
  guided and predictive smooth pursuit dynamics},\ }\href@noop {} {\bibfield
  {journal} {\bibinfo  {journal} {Journal of neuroscience}\ }\textbf {\bibinfo
  {volume} {33}},\ \bibinfo {pages} {17301} (\bibinfo {year}
  {2013})}\BibitemShut {NoStop}%
\bibitem [{\citenamefont {Meinhold}\ and\ \citenamefont
  {Singpurwalla}(1983)}]{MS83}%
  \BibitemOpen
  \bibfield  {author} {\bibinfo {author} {\bibfnamefont {R.~J.}\ \bibnamefont
  {Meinhold}}\ and\ \bibinfo {author} {\bibfnamefont {N.~D.}\ \bibnamefont
  {Singpurwalla}},\ }\bibfield  {title} {\bibinfo {title} {Understanding the
  kalman filter},\ }\href@noop {} {\bibfield  {journal} {\bibinfo  {journal}
  {The American Statistician}\ }\textbf {\bibinfo {volume} {37}},\ \bibinfo
  {pages} {123} (\bibinfo {year} {1983})}\BibitemShut {NoStop}%
\bibitem [{\citenamefont {Rao}(1999)}]{R99}%
  \BibitemOpen
  \bibfield  {author} {\bibinfo {author} {\bibfnamefont {R.~P.}\ \bibnamefont
  {Rao}},\ }\bibfield  {title} {\bibinfo {title} {An optimal estimation
  approach to visual perception and learning},\ }\href@noop {} {\bibfield
  {journal} {\bibinfo  {journal} {Vision research}\ }\textbf {\bibinfo {volume}
  {39}},\ \bibinfo {pages} {1963} (\bibinfo {year} {1999})}\BibitemShut
  {NoStop}%
\bibitem [{\citenamefont {Lewis}\ \emph {et~al.}(2008)\citenamefont {Lewis},
  \citenamefont {Xie},\ and\ \citenamefont {Popa}}]{lewis2008}%
  \BibitemOpen
  \bibfield  {author} {\bibinfo {author} {\bibfnamefont {F.}~\bibnamefont
  {Lewis}}, \bibinfo {author} {\bibfnamefont {L.}~\bibnamefont {Xie}},\ and\
  \bibinfo {author} {\bibfnamefont {D.}~\bibnamefont {Popa}},\ }\href@noop {}
  {\emph {\bibinfo {title} {Optimal and Robust Estimation}}},\ \bibinfo
  {edition} {2nd}\ ed.\ (\bibinfo  {publisher} {CRC Press},\ \bibinfo {year}
  {2008})\BibitemShut {NoStop}%
\bibitem [{\citenamefont {Dixit}(2020)}]{D20}%
  \BibitemOpen
  \bibfield  {author} {\bibinfo {author} {\bibfnamefont {P.~D.}\ \bibnamefont
  {Dixit}},\ }\bibfield  {title} {\bibinfo {title} {Thermodynamic inference of
  data manifolds},\ }\href@noop {} {\bibfield  {journal} {\bibinfo  {journal}
  {Physical Review Research}\ }\textbf {\bibinfo {volume} {2}},\ \bibinfo
  {pages} {023201} (\bibinfo {year} {2020})}\BibitemShut {NoStop}%
\bibitem [{\citenamefont {Gampe}\ and\ \citenamefont {Griffin}(2023)}]{GG23}%
  \BibitemOpen
  \bibfield  {author} {\bibinfo {author} {\bibfnamefont {H.}~\bibnamefont
  {Gampe}}\ and\ \bibinfo {author} {\bibfnamefont {C.}~\bibnamefont
  {Griffin}},\ }\bibfield  {title} {\bibinfo {title} {Dynamics of a binary
  option market with exogenous information and price sensitivity},\ }\href@noop
  {} {\bibfield  {journal} {\bibinfo  {journal} {Communications in Nonlinear
  Science and Numerical Simulation}\ }\textbf {\bibinfo {volume} {118}},\
  \bibinfo {pages} {106994} (\bibinfo {year} {2023})}\BibitemShut {NoStop}%
\bibitem [{\citenamefont {Miller}\ \emph {et~al.}(2020)\citenamefont {Miller},
  \citenamefont {Xiang},\ and\ \citenamefont {Kesidis}}]{MXK20}%
  \BibitemOpen
  \bibfield  {author} {\bibinfo {author} {\bibfnamefont {D.~J.}\ \bibnamefont
  {Miller}}, \bibinfo {author} {\bibfnamefont {Z.}~\bibnamefont {Xiang}},\ and\
  \bibinfo {author} {\bibfnamefont {G.}~\bibnamefont {Kesidis}},\ }\bibfield
  {title} {\bibinfo {title} {Adversarial learning targeting deep neural network
  classification: A comprehensive review of defenses against attacks},\
  }\href@noop {} {\bibfield  {journal} {\bibinfo  {journal} {Proceedings of the
  IEEE}\ }\textbf {\bibinfo {volume} {108}},\ \bibinfo {pages} {402} (\bibinfo
  {year} {2020})}\BibitemShut {NoStop}%
\bibitem [{\citenamefont {Verdonck}\ \emph {et~al.}(2021)\citenamefont
  {Verdonck}, \citenamefont {Loossens},\ and\ \citenamefont
  {Philiastides}}]{VLP21}%
  \BibitemOpen
  \bibfield  {author} {\bibinfo {author} {\bibfnamefont {S.}~\bibnamefont
  {Verdonck}}, \bibinfo {author} {\bibfnamefont {T.}~\bibnamefont {Loossens}},\
  and\ \bibinfo {author} {\bibfnamefont {M.~G.}\ \bibnamefont {Philiastides}},\
  }\bibfield  {title} {\bibinfo {title} {The leaky integrating threshold and
  its impact on evidence accumulation models of choice response time (rt).},\
  }\href@noop {} {\bibfield  {journal} {\bibinfo  {journal} {Psychological
  Review}\ }\textbf {\bibinfo {volume} {128}},\ \bibinfo {pages} {203}
  (\bibinfo {year} {2021})}\BibitemShut {NoStop}%
\bibitem [{\citenamefont {Nickerson}(1998)}]{N98}%
  \BibitemOpen
  \bibfield  {author} {\bibinfo {author} {\bibfnamefont {R.~S.}\ \bibnamefont
  {Nickerson}},\ }\bibfield  {title} {\bibinfo {title} {Confirmation bias: A
  ubiquitous phenomenon in many guises},\ }\href@noop {} {\bibfield  {journal}
  {\bibinfo  {journal} {Review of general psychology}\ }\textbf {\bibinfo
  {volume} {2}},\ \bibinfo {pages} {175} (\bibinfo {year} {1998})}\BibitemShut
  {NoStop}%
\bibitem [{\citenamefont {Hofbauer}(1996)}]{H96}%
  \BibitemOpen
  \bibfield  {author} {\bibinfo {author} {\bibfnamefont {J.}~\bibnamefont
  {Hofbauer}},\ }\bibfield  {title} {\bibinfo {title} {Evolutionary dynamics
  for bimatrix games: A {H}amiltonian system?},\ }\href@noop {} {\bibfield
  {journal} {\bibinfo  {journal} {J. Math. Bio}\ }\textbf {\bibinfo {volume}
  {34}},\ \bibinfo {pages} {675} (\bibinfo {year} {1996})}\BibitemShut
  {NoStop}%
\bibitem [{\citenamefont {Mangasarian}(1966)}]{Mang66}%
  \BibitemOpen
  \bibfield  {author} {\bibinfo {author} {\bibfnamefont {O.~L.}\ \bibnamefont
  {Mangasarian}},\ }\bibfield  {title} {\bibinfo {title} {Sufficient conditions
  for the optimal control of nonlinear systems},\ }\href@noop {} {\bibfield
  {journal} {\bibinfo  {journal} {SIAM J. Control}\ }\textbf {\bibinfo {volume}
  {4}},\ \bibinfo {pages} {139} (\bibinfo {year} {1966})}\BibitemShut {NoStop}%
\bibitem [{\citenamefont {Friesz}(2010)}]{Friesz10}%
  \BibitemOpen
  \bibfield  {author} {\bibinfo {author} {\bibfnamefont {T.}~\bibnamefont
  {Friesz}},\ }\href@noop {} {\emph {\bibinfo {title} {{Dynamic Optimization
  and Differential Games}}}}\ (\bibinfo  {publisher} {Springer},\ \bibinfo
  {year} {2010})\BibitemShut {NoStop}%
\end{thebibliography}%

\end{document}